\documentclass[review, authoryear]{elsarticle}
\usepackage{amsmath}
\usepackage{booktabs}
\usepackage{multirow}
\usepackage{subfig}
\usepackage{color}

\graphicspath{{figures/}}

\journal{Elsevier}
\newcommand{\diff}{\,\mathrm{d}}

\begin{document}

\begin{frontmatter}
\title{Parallel finite volume simulation of the spherical shell dynamo with pseudo-vacuum magnetic boundary conditions}

\author[must,ucas]{Liang Yin}
\ead{lyin@must.edu.mo}

\author[pku]{Chao Yang\corref{cor}}
\ead{chao\_yang@pku.edu.cn}

\author[ucas]{Shi-Zhuang Ma}
\ead{szma@ucas.edu.cn}

\author[iapcm]{Ying Cai}
\ead{cai\_ying@iapcm.ac.cn}

\author[must]{Keke Zhang}
\ead{K.Zhang@exeter.ac.uk}

\cortext[cor]{Corresponding author}

\address[must]{State Key Laboratory of Lunar and Planetary Sciences, Macau University of Science and Technology, Macau, China}
\address[ucas]{School of Engineering Science, University of Chinese Academy of Sciences, Beijing 100049, China}
\address[pku]{School of Mathematical Sciences, Peking University, Beijing 100871, China}
\address[iapcm]{Institute of Applied Physics and Computational Mathematics, Beijing 100094, China}

\begin{abstract}
In this paper, we study the parallel simulation of the magnetohydrodynamic (MHD) dynamo in a rapidly rotating spherical shell with pseudo-vacuum magnetic boundary conditions.
A second-order finite volume scheme based on a collocated quasi-uniform cubed-sphere grid is applied to the spatial discretization of the MHD dynamo equations.
To ensure the solenoidal condition of the magnetic field, we adopt a widely-used approach whereby a pseudo-pressure is introduced into the induction equation.
The temporal integration is split by a second-order approximate factorization approach, resulting in two linear algebraic systems both solved by a preconditioned Krylov subspace iterative method.
A multi-level restricted additive Schwarz preconditioner based on domain decomposition and multigrid method is then designed to improve the efficiency and scalability.
Accurate numerical solutions of two benchmark cases are obtained with our code, comparable to the existing local method results.
Several large-scale tests performed on the Sunway TaihuLight supercomputer show good strong and weak scalabilities and a noticeable improvement from the multi-level preconditioner with up to 10368 processor cores.

\end{abstract}


\end{frontmatter}

\section{Introduction}

The magnetic field of the Earth as well as many other planets is widely thought to be generated by the convection of the electrically conducting fluid in the outer core, which creates the so-called self-consistent dynamo action \citep{Zhang2000, Christensen2007, Wicht2010, Jones2011, Roberts2013, Moffatt2019, Deguen2020}.
Due to a series of reasons \citep{Aurnou2015, Vantieghem2016}, it is a challenging task to fully understand the dynamics in planetary fluid cores.
Starting with the pioneering work by \citet{Glatzmaier1995}, \citet{Kageyama1995} and \citet{Kuang1997}, significant progresses have been made in understanding of the origin and evolution of the Earth's magnetic field by means of numerical dynamo simulations \citep{Sheyko2016, Christensen2018, Petitdemange2018, Aubert2019}.
Although there has been a number of numerical codes for dynamo modelling developed independently by various groups \citep{Matsui2016}, there is still a long way from achieving dynamo simulations with physically realistic parameters, due mainly to the difficulties in extreme-scale spatial resolutions \citep{Glatzmaier2002, Jones2011, Roberts2013, Aurnou2015}.
Tremendous amounts of computing resources are required for such extreme-resolution dynamo simulations, which can only be made possible with the aid of massively parallel supercomputers \citep{Harder2005, Chan2006}.
Innovations in the numerical algorithms and their applications on massively parallel supercomputers are likely beneficial to extend the parameter regime in dynamo simulations towards more realistic values relevant to the planetary cores \citep{Matsui2002, Matsui2004b, Harder2005, Chan2006, Chan2007, Matsui2016}.

As reported in \citet{Matsui2016}, a majority of the existing widely-used numerical dynamo models employ global-nature spectral methods, which are based on the poloidal-toroidal decomposition and spherical harmonic expansions.
The solenoidal condition of the velocity and magnetic field and the insulating boundary condition for the magnetic field can be easily dealt with in such methods.
However, a significant number of global communications are usually required for the computation of nonlinear terms, which could make spectral methods less suitable for large-scale parallel computations \citep{Harder2005, Chan2006, Chan2007, Wicht2009}.
Besides, spectral methods are often hard to be adapted to more complicated domains without a spherical symmetry \citep{Iskakov2004, Vantieghem2016} and local horizontal variations of physical properties such as the electrical conductivity \citep{Chan2007, Wicht2009}.
In contrast, local discretization approaches, such as finite volume and finite element methods, show more potentials for parallel computations and could be more flexible to complicated domains and local variations, thus are bringing an increasing interest in dynamo simulations \citep[e.g.][]{Kageyama1995, Kageyama1997, Hejda2003, Hejda2004, Kageyama2005, Harder2005, Matsui2002, Matsui2004b, Matsui2004a, Matsui2005, Chan2001b, Chan2001a, Chan2006, Chan2007, Vantieghem2016, Yin2017, Yin2019}.
However, the applications of the local methods in dynamo simulations still face several difficulties, such as: (i) the solenoidal conditions of the velocity and magnetic field, (ii) the insulating boundary condition for the magnetic field, and (iii) parallel scalability.
To cope with the solenoidal constraint, most of the dynamo simulations adopt a projection-based method introduced by \citet{Toth2000}, in which a pseudo-pressure gradient is added into the induction equation and the pseudo-pressure is interpreted as an effecting projection of the provisional magnetic field onto the solenoidal space, just as the pressure in momentum equation.
This method has been applied successfully to a large number of dynamo models \citep[e.g.][]{Chan2001b, Harder2005, Chan2007, Vantieghem2016}.

The exterior space outside the Earth's core is generally thought as an electrically insulating medium, resulting in a non-local nature of the magnetic boundary condition since the solution of a Laplace equation for the magnetic scalar potential in the infinite exterior domain is theoretically required.
This insulating boundary condition brings substantial difficulties in dynamo simulations with local methods.
A straightforward approximation is to replace the infinite exterior extent with a finite domain since the magnetic scalar potential declines as $O(r^{-2})$ in the insulating exterior and can be approximated as zero at the location far enough from the fluid domain of interest.
Based on this approximation, several dynamo models with local methods treat the insulating boundary condition via an extra numerical cost in a finite exterior domain. 
Examples include \citet{Chan2001b} and \citet{Chan2007} in which a weak conductivity approximation is introduced and \citet{Matsui2005} where a formulation of the magnetic vector potential is used.
Another approach is to introduce a mathematically equivalent boundary integral formulation proposed by \citet{Iskakov2004}, in which the 3-D Laplace equation for the magnetic potential is recast as a 2-D integral equation on the boundary surface next to the insulating medium.
However, such boundary integral approach usually leads to a higher computational cost due to the dense coefficient matrix and the global communication between all processors handling the boundary points.
On the other hand, pseudo-vacuum boundary conditions, which prescribe the tangential component of the magnetic field on the boundary to be zero, since first adopted by \citet{Kageyama1995} in their finite difference code, have become a popular alternative.
Implementation of such conditions in local methods is straightforward and no extra numerical cost or global communication is required.
Though these conditions may result in quite different numerical solutions from the insulating condition \citep{Harder2005, Jackson2014}, it may be quite suitable to apply them to the benchmark studies, as was done in \citet{Jackson2014} and \citet{Vantieghem2016}, thus providing a convenient way to validate dynamo codes with local methods.

The parallel scalability, which is the key limiting factor for the massive-scale simulations, is frequently investigated in dynamo simulations to different degrees \citep[e.g.][]{Harder2005, Chan2006, Chan2007, Vantieghem2016}.
In particular, parallel performance benchmarks from 15 widely-used parallel dynamo models are thoroughly reported by \citet{Matsui2016} but only two codes based on local methods are included therein.
While local methods show great potentials for the massive-scale dynamo simulations, the study on improving the parallel scalability of local codes is still of critical importance.
The temporal integration in the projection-based local models generally involves a fractional step algorithm consisting of a prediction procedure and a correction step.
In practice, most of the computing time in the temporal integration is spent on the numerical solution of the pressure Poisson equation in the correction step \citep{Harder2005, Vantieghem2016, Yin2017}.
To improve the parallel performance of this type of code, the design of a scalable solver for the prediction equation and especially for the pressure Poisson equation is highly desirable.

In this work, we present a parallel finite volume solution for the convection-driven magnetohydrodynamic (MHD) dynamo in a rapidly rotating spherical shell with pseudo-vacuum magnetic boundary conditions.
As a continuation to our previous work \citep{Yin2017} on the non-magnetic convection problem, this paper inherits the approximate factorization method in temporal integration and the finite volume scheme on a collocated quasi-uniform cubed-sphere grid, and focuses on the algorithms and implementations related to the magnetic field.
In addition to that, efforts have also been made on the design of a multi-level restricted additive Schwarz preconditioner based on domain decomposition and multigrid method to improve the efficiency and scalability.
It is worth mentioning that the adopted cubed-sphere grid can be easily extended to ellipsoidal shell domains and, in theory, other geometries which can be expressed by a similar projection relationship.
Besides, a non-staggered finite volume scheme in general curvilinear geometries has been successfully developed for the compressible MHD equations \citep{Chacon2004} and an implicit Newton-Krylov solver with a well-designed preconditioner \citep{Chacon2008} has shown the good efficiency of the multigrid method in such problem.
In contrast, this work specifically concerns the finite volume discretization of the incompressible MHD dynamo equations on the composite cubed-sphere grid and the efficiency of the semi-implicit fractional step solver for the dynamo problem.

The remainder of the paper is organized as follows.
We first present the governing equations for the MHD dynamo problem and the boundary conditions in Section \ref{mathematicalmodel}.
Then in Section \ref{numericalmethods}, we introduce the numerical methods including the temporal integration, spatial discretization and parallel multi-level solver.
The numerical results about the validation and parallel performance are reported in Section \ref{numericalresults}.
We conclude the paper in Section \ref{conclusions}.

\section{Mathematical model}
\label{mathematicalmodel}

In this work we focus our discussion on solving the convection-driven MHD dynamo problem in a rapidly rotating spherical shell with pseudo-vacuum magnetic boundary conditions.
The spherical shell, with outer radius $r_o$ and inner radius $r_i$, is filled with electrically conducting viscous incompressible fluids and rotates with a constant angular velocity $\mathbf{\Omega}=\Omega \hat{z}$, where $\hat{z}$ is a unit vector parallel to the axis of rotation.
The incompressible fluids in the spherical shell is assumed to satisfy the Boussinesq approximation, with the density $\rho$, kinetic viscosity $\nu$, thermal diffusivity $\kappa$, thermal expand coefficiency $\alpha$, magnetic diffusivity $\eta$, magnetic permeability $\mu$.
The temperatures on the inner and outer boundaries are fixed to be $T_i$ and $T_o$, respectively and the temperature difference is denoted by $\Delta T = T_i - T_o$.
Choosing the shell thickness $D = r_o - r_i$ as the fundamental length scale, $D^2/\nu$ as the time scale, $\nu/D$ as the velocity scale, $\Delta T$ as the temperature scale, $\sqrt{\rho \mu \eta \Omega}$ as the magnetic field scale, $\rho \nu \Omega$ as the pressure scale, we can obtain the non-dimensional governing equations of the MHD dynamo problem
\begin{gather}
	E \left( \frac{\partial \mathbf{u}}{\partial t} + \mathbf{u} \cdot \nabla \mathbf{u} - \nabla^2 \mathbf{u} \right) + 2 \hat{z} \times \mathbf{u} + \nabla P = Ra \frac{\mathbf{r}}{r_o} \left( T - T_0 \right) + \frac{1}{Pm} \mathbf{B} \cdot \nabla \mathbf{B}, \label{eqn:momentum}\\
	\nabla \cdot \mathbf{u} = 0, \\
	\frac{\partial T}{\partial t} + \mathbf{u} \cdot \nabla T = \frac{1}{Pr} \nabla^2 T, \\
	\frac{\partial \mathbf{B}}{\partial t} = \nabla \times \left( \mathbf{u} \times \mathbf{B} \right) + \frac{1}{Pm} \nabla^2 \mathbf{B}, \label{eqn:induction}\\
	\nabla \cdot \mathbf{B} = 0, \label{eqn:magdivergence}
\end{gather}
where $\mathbf{u}$, $P$, $\mathbf{B}$, $T$, $T_0$ and $\mathbf{r}$ are non-dimensional velocity, reduced pressure, magnetic field, temperature, reference temperature and spatial position vector, respectively.
The reference temperature $T_0$ can be expressed by
\begin{equation}
T_0(r) = r_i \left( \frac{r_o}{r} - 1 \right),
\end{equation}
where the dimensionless radii are set to be $r_i = 7/13, r_o = 20/13$.
The non-dimensional parameters $E, Ra, Pr, Pm$ in the above equations are Ekman number, modified Rayleigh number, Prandtl number and magnetic Prandtl number respectively and defined by 
\begin{equation}
	E = \frac{\nu}{\Omega D^2}, \quad Ra = \frac{\alpha g_o \Delta T D}{\nu \Omega}, \quad Pr = \frac{\nu}{\kappa}, \quad Pm = \frac{\nu}{\eta},
\end{equation}
where $g_o$ is the gravitational acceleration at the outer radius.

An equivalent form of the magnetic induction equation \eqref{eqn:induction} is 
\begin{equation}
	\frac{\partial \mathbf{B}}{\partial t} = \nabla \times \left( \mathbf{u} \times \mathbf{B} \right) - \frac{1}{Pm} \nabla \times \nabla \times \mathbf{B}.
\end{equation}
Applying the divergence operator to the above equation, we can obtain
\begin{equation}
	\frac{\partial \left( \nabla \cdot \mathbf{B} \right)}{\partial t} = 0.
\end{equation}
This equation indicates that the magnetic field will keep the divergence-free constraint \eqref{eqn:magdivergence} all the time in the evolution if the initial magnetic field is solenoidal.
In numerical simulations, however, the divergence-free constraint is difficult to maintain.
To overcome this difficulty, we adopt a technique of introducing a pseudo-pressure gradient into the magnetic induction equation \citep{Toth2000, Harder2005, Chan2007, Vantieghem2016} and replace the induction equation \eqref{eqn:induction} with the following equation
\begin{equation}
	\frac{\partial \mathbf{B}}{\partial t} + \frac 1E \nabla P_b = \nabla \times \left( \mathbf{u} \times \mathbf{B} \right) + \frac{1}{Pm} \nabla^2 \mathbf{B}, \label{eqn:pinduction}
\end{equation}
where $P_b$ is the pseudo-pressure.
Therefore, a projection method similar to the well-known treatment \citep{Chorin1968, Guermond2006} of velocity fields can be applied to the magnetic field to ensure the divergence-free constraint.

Replacing the temperature $T$ with an auxiliary temperature variable $\Theta = T - T_0$, we can rewrite the non-dimensional governing equations as
\begin{gather}
	\frac{\partial \mathbf{u}}{\partial t} - \nabla^2 \mathbf{u} + \frac 2E \hat{z} \times \mathbf{u} - \frac{Ra}{E r_o} \Theta \mathbf{r} + \frac 1E \nabla P = - \mathbf{u} \cdot \nabla \mathbf{u} + \frac{1}{E Pm} \mathbf{B} \cdot \nabla \mathbf{B}, \label{eqn:velocity}\\
	\frac{\partial \Theta}{\partial t} - \frac{1}{Pr} \nabla^2 \Theta -\frac{r_i r_o}{r^3} \mathbf{u} \cdot \mathbf{r} = - \mathbf{u} \cdot \nabla \Theta, \\
	\frac{\partial \mathbf{B}}{\partial t} - \frac{1}{Pm} \nabla^2 \mathbf{B} + \frac 1E \nabla P_b = \left( \mathbf{B} \cdot \nabla \right) \mathbf{u} - \left( \mathbf{u} \cdot \nabla \right) \mathbf{B}, \\
	\nabla \cdot \mathbf{u} = 0, \\
	\nabla \cdot \mathbf{B} = 0. \label{eqn:divb}
\end{gather}
To solve the above system, it is necessary to apply proper boundary conditions to the velocity, temperature and magnetic field.
On the shell boundaries, we employ the no-slip condition for the velocity and isothermal condition for the temperature,
\begin{equation}
	\mathbf{u} = 0, \quad \Theta = 0, \qquad r = r_i, r_o. \label{eqn:utb}
\end{equation}
And for the magnetic field, we adopt the pseudo-vacuum boundary condition
\begin{equation}
	\mathbf{r} \times \mathbf{B} = 0, \qquad r = r_i, r_o, \label{eqn:pvc}
\end{equation}
on the shell boundaries, which indicates that the tangential component of $\mathbf{B}$ is zero and only the normal component exists \citep{Kageyama1995, Jackson2014}.
The value of the normal component on the boundaries can be constrained by the solenoidal condition $\nabla \cdot \mathbf{B} = 0$.

\section{Numerical methods}
\label{numericalmethods}

In this section, we present the proposed numerical methods to discretize and solve the MHD dynamo equations \eqref{eqn:velocity}--\eqref{eqn:divb} with the boundary conditions \eqref{eqn:utb}--\eqref{eqn:pvc}.
Since some of the algorithms have already been introduced in a previous work that does not involve the magnetic field \citep{Yin2017}, we will focus on the treatments of the issues related to the magnetic field.

\subsection{Temporal integration scheme}
A second-order approximate factorization method \citep{Dukowicz1992} was applied successfully to deal with the temporal integration and ensure the solenoidal condition of the velocity in the non-magnetic convection problem \citep{Chan2006, Yin2017}.
In this section, we inherit the approximate factorization method and extend it to the temporal integration of the magnetic field.
The dynamo governing equations \eqref{eqn:velocity}--\eqref{eqn:divb} can be rewritten in the following operator form
\begin{gather}
	\frac{\partial \mathbf{u}}{\partial t} - L_1(\mathbf{u}) - R_1(\Theta) + G(P) = \mathbf{f}_1, \label{eqn:operatorvelocity}\\
	\frac{\partial \Theta}{\partial t} - L_2(\Theta) - R_2(\mathbf{u}) = f_2, \\
	\frac{\partial \mathbf{B}}{\partial t} - L_3(\mathbf{B}) + G(P_b) = \mathbf{f}_3, \\
	D(\mathbf{u}) = 0, \\
	D(\mathbf{B}) = 0, \label{eqn:operatordivb}
\end{gather}
where the $L_1$, $R_1$, $G$, $L_2$, $R_2$, $L_3$ and $D$ are linear operators defined by
\begin{equation}
	\label{eqn:linearoperator}
	\left\{
	\begin{aligned}
		&L_1(\mathbf{u}) = \nabla^2 \mathbf{u} - \frac 2E (\hat{z} \times \mathbf{u}), \quad R_1(\Theta) = \frac{Ra}{E r_o} (\Theta \mathbf{r}), \quad G(P) = \frac 1E \nabla P, \\
		&L_2(\Theta) = \frac{1}{Pr} \nabla^2 \Theta, \quad R_2(\mathbf{u}) = \frac{r_i r_o}{r^3} \mathbf{u} \cdot \mathbf{r}, \quad L_3(\mathbf{B}) = \frac{1}{Pm}\nabla^2 \mathbf{B}, \\
		&G(P_b) = \frac 1E \nabla P_b, \quad D(\mathbf{u}) = \nabla \cdot \mathbf{u}, \quad D(\mathbf{B}) = \nabla \cdot \mathbf{B},
	\end{aligned}
	\right.
\end{equation}
and the right hand sides are nonlinear terms
\begin{equation}
	\label{eqn:nonlinearoperator}
	\left\{
	\begin{aligned}
		&\mathbf{f}_1 = -\mathbf{u} \cdot \nabla \mathbf{u} + \frac{1}{E Pm} \mathbf{B} \cdot \nabla \mathbf{B}, \\
		&f_2 = -\mathbf{u} \cdot \nabla \Theta, \\
		&\mathbf{f}_3 = \mathbf{B} \cdot \nabla \mathbf{u} - \mathbf{u} \cdot \nabla \mathbf{B}.
	\end{aligned}
	\right.
\end{equation}

Applying the Crank-Nicolson scheme to the linear operators and discretizing all terms spatially, the governing equations \eqref{eqn:operatorvelocity}--\eqref{eqn:operatordivb} can be fully discretized as
\begin{align}
	&\frac{\mathbf{u}^{n+1} - \mathbf{u}^n}{\Delta t} - \mathcal{L}_1 \left( \frac{\mathbf{u}^{n+1} + \mathbf{u}^n}{2} \right) - \mathcal{R}_1 \left( \frac{\Theta^{n+1} + \Theta^n}{2} \right) + \mathcal{G} \left( \frac{P^{n+1} + P^n}{2} \right) \notag\\
	&= \hat{\mathbf{f}}_1^{n+\frac 12} + O \left( \Delta t^2 \right), \label{eqn:velocitycn}\\
	&\frac{\Theta^{n+1} - \Theta^n}{\Delta t} - \mathcal{L}_2 \left( \frac{\Theta^{n+1} + \Theta^n}{2} \right) - \mathcal{R}_2 \left( \frac{\mathbf{u}^{n+1} + \mathbf{u}^n}{2} \right) \notag\\
	&= \hat{f}_2^{n+\frac 12} + O \left( \Delta t^2 \right), \\
	&\frac{\mathbf{B}^{n+1} - \mathbf{B}^n}{\Delta t} - \mathcal{L}_3 \left( \frac{\mathbf{B}^{n+1} + \mathbf{B}^n}{2} \right) + \mathcal{G} \left( \frac{P_b^{n+1} + P_b^n}{2} \right) \notag\\
	&= \hat{\mathbf{f}}_3^{n+\frac 12} + O \left( \Delta t^2 \right), \label{eqn:magcn}\\
	&\mathcal{D} \left( \mathbf{u}^{n+1} \right) = \mathcal{D} \left( \mathbf{B}^{n+1} \right) = 0, \label{eqn:divtime}
\end{align}
where $\Delta t$ is the time step size and $n$ denotes the time step number.
$\mathcal{L}_1$, $\mathcal{R}_1$, $\mathcal{G}$, $\mathcal{L}_2$, $\mathcal{R}_2$, $\mathcal{L}_3$ and $\mathcal{D}$ are discrete linear operators corresponding to the linear terms in equation \eqref{eqn:linearoperator}.
Note that $\mathcal{L}_1$ is not a symmetric operator due to the implicit treatment of the Coriolis force term.
$\hat{\mathbf{f}}_1$, $\hat{f}_2$ and $\hat{\mathbf{f}}_3$ are the discrete nonlinear terms corresponding to equation \eqref{eqn:nonlinearoperator}.
The nonlinear terms $\hat{\mathbf{f}}_1$, $\hat{f}_2$ and $\hat{\mathbf{f}}_3$ are calculated by the second-order Adams-Bashforth formula
\begin{equation}
	f^{n+\frac 12} = \frac 32 f^n - \frac 12 f^{n-1} + O \left( \Delta t^2 \right),
\end{equation}
except the first time step by a first-order approximation $f^{\frac 12} = f^0$.
The spatial discretization schemes of the linear and nonlinear terms in the above equations will be described in Section \ref{spatialdiscretization}.

Moving the unknown terms about the time step $n+1$ to the left hand sides and others to the right, the equations \eqref{eqn:velocitycn}--\eqref{eqn:magcn} can be transformed into the following form
\begin{align}
	&\tilde{\mathbf{u}}^{n+1} - \frac{\Delta t}{2} \mathcal{L}_1 \left( \tilde{\mathbf{u}}^{n+1} \right) - \frac{\Delta t}{2} \mathcal{R}_1 \left( \Theta^{n+1} \right) 
	= \hat{\mathbf{u}}^n + \frac{\Delta t}{2} \mathcal{L}_1 \left( \hat{\mathbf{u}}^n \right) + \frac{\Delta t}{2} \mathcal{R}_1 \left( \Theta^n \right) \notag \\
	& + \Delta t \hat{\mathbf{f}}_1^{n+\frac 12} - \frac{\Delta t^2}{4} \mathcal{L}_1 \mathcal{G} \left( P^{n+1} - P^n \right) + O \left( \Delta t^3 \right), \\
	& \Theta^{n+1} - \frac{\Delta t}{2} \mathcal{L}_2 \left( \Theta^{n+1} \right) - \frac{\Delta t}{2} \mathcal{R}_2 \left( \tilde{\mathbf{u}}^{n+1} \right) 
	= \Theta^n + \frac{\Delta t}{2} \mathcal{L}_2 \left( \Theta^n \right) + \frac{\Delta t}{2} \mathcal{R}_2 \left( \hat{\mathbf{u}}^n \right) \notag \\
	& + \Delta t \hat{f}_2^{n+\frac 12} - \frac{\Delta t^2}{4} \mathcal{R}_2 \mathcal{G} \left( P^{n+1} - P^n \right) + O \left( \Delta t^3 \right), \\
	& \tilde{\mathbf{B}}^{n+1} - \frac{\Delta t}{2} \mathcal{L}_3 \left( \tilde{\mathbf{B}}^{n+1} \right) 
	= \hat{\mathbf{B}}^n + \frac{\Delta t}{2} \mathcal{L}_3 \left( \hat{\mathbf{B}}^n \right) \notag \\
	& + \Delta t \hat{\mathbf{f}}_3^{n+\frac 12} - \frac{\Delta t^2}{4} \mathcal{L}_3 \mathcal{G} \left( P_b^{n+1} - P_b^n \right) + O \left( \Delta t^3 \right),
\end{align}
where $\tilde{\mathbf{u}}^{n+1}$, $\hat{\mathbf{u}}^n$, $\tilde{\mathbf{B}}^{n+1}$ and $\hat{\mathbf{B}}^n$ are intermediate velocities and magnetic fields defined by
\begin{gather}
	\tilde{\mathbf{u}}^{n+1} = \mathbf{u}^{n+1} + \frac{\Delta t}{2} \mathcal{G} \left( P^{n+1} \right), \quad 
	\tilde{\mathbf{B}}^{n+1} = \mathbf{B}^{n+1} + \frac{\Delta t}{2} \mathcal{G} \left( P_b^{n+1} \right), \label{eqn:ubtilde} \\
	\label{eqn:ubhat}
	\hat{\mathbf{u}}^n = \mathbf{u}^n - \frac{\Delta t}{2} \mathcal{G} \left( P^n \right), \quad
	\hat{\mathbf{B}}^n = \mathbf{B}^n - \frac{\Delta t}{2} \mathcal{G} \left( P_b^n \right).
\end{gather}

By expanding $P^{n+1}$ in Taylor series about $P^n$, It can be observed that the pressure terms $(- {\Delta t^2}/{4}) \mathcal{L}_1 \mathcal{G} ( P^{n+1} - P^n )$, $(- {\Delta t^2}/{4}) \mathcal{R}_2 \mathcal{G} ( P^{n+1} - P^n )$ and $(- {\Delta t^2}/{4}) \mathcal{L}_3 \mathcal{G} ( P_b^{n+1} - P_b^n )$ are all of $O(\Delta t^3)$ .
Discarding these terms as well as the temporal truncation error, we can obtain the following fully discretized time stepping equations for the intermediate velocity $\tilde{\mathbf{u}}^{n+1}$, temperature $\Theta$ and intermediate magnetic field $\tilde{\mathbf{B}}^{n+1}$
\begin{align}
	\label{eqn:tsvelocity}
	&\tilde{\mathbf{u}}^{n+1} - \frac{\Delta t}{2} \mathcal{L}_1 \left( \tilde{\mathbf{u}}^{n+1} \right) - \frac{\Delta t}{2} \mathcal{R}_1 \left( \Theta^{n+1} \right) 
	= \hat{\mathbf{u}}^n + \frac{\Delta t}{2} \mathcal{L}_1 \left( \hat{\mathbf{u}}^n \right) + \frac{\Delta t}{2} \mathcal{R}_1 \left( \Theta^n \right) \notag \\
	& + \Delta t \hat{\mathbf{f}}_1^{n+\frac 12} , \\
	\label{eqn:tstemperature}
	& \Theta^{n+1} - \frac{\Delta t}{2} \mathcal{L}_2 \left( \Theta^{n+1} \right) - \frac{\Delta t}{2} \mathcal{R}_2 \left( \tilde{\mathbf{u}}^{n+1} \right) 
	= \Theta^n + \frac{\Delta t}{2} \mathcal{L}_2 \left( \Theta^n \right) + \frac{\Delta t}{2} \mathcal{R}_2 \left( \hat{\mathbf{u}}^n \right) \notag \\
	& + \Delta t \hat{f}_2^{n+\frac 12} , \\
	\label{eqn:tsmag}
	& \tilde{\mathbf{B}}^{n+1} - \frac{\Delta t}{2} \mathcal{L}_3 \left( \tilde{\mathbf{B}}^{n+1} \right) 
	= \hat{\mathbf{B}}^n + \frac{\Delta t}{2} \mathcal{L}_3 \left( \hat{\mathbf{B}}^n \right) + \Delta t \hat{\mathbf{f}}_3^{n+\frac 12} .
\end{align}
Applying the divergence operator to equation \eqref{eqn:ubtilde} and subtracting equation \eqref{eqn:divtime}, we can obtain two Poisson equations for the pressure $P^{n+1}$ and pseudo-pressure $P_b^{n+1}$ respectively
\begin{gather}
	\label{eqn:tsp}
	\frac{\Delta t}{2} \mathcal{D} \mathcal{G} \left( P^{n+1} \right) = \mathcal{D} \left( \tilde{\mathbf{u}}^{n+1} \right), \\
	\label{eqn:tspb}
	\frac{\Delta t}{2} \mathcal{D} \mathcal{G} \left( P_b^{n+1} \right) = \mathcal{D} \left( \tilde{\mathbf{B}}^{n+1} \right).
\end{gather}
Boundary conditions for the pressure $P$ and pseudo-pressure $P_b$ are required to solve the above equations.
The Neumann boundary condition $\mathbf{n} \cdot \nabla P = 0$ is applied to the pressure $P$, where $\mathbf{n}$ denotes the outward normal unit vector, and a detailed discussion can be found in \citet{Yin2017}.
The pseudo-vacuum condition \eqref{eqn:pvc} is applied to both $P$ and $P_b$, and according to equation \eqref{eqn:ubtilde}, we can obtain $\mathbf{n} \times \nabla P_b = 0$, which indicates a Dirichlet boundary condition of $P_b$ on the boundaries.

As a result, two linear algebraic systems are obtained including the equations for the velocity, temperature and magnetic field (VTBE) \eqref{eqn:tsvelocity}--\eqref{eqn:tsmag} and the equations for the pressure and pseudo-pressure (PPBE) \eqref{eqn:tsp}--\eqref{eqn:tspb}.
Based on these two linear systems, a predictor-corrector procedure is adopted to obtain the required numerical solutions.
The outline of the resulting semi-implicit time stepping scheme can be summarized as follows: 
\begin{description}
	\item[Step 1:] According to the previous values $\mathbf{u}^n$, $\Theta^n$, $\mathbf{B}^n$ and $P^n$, calculate $\hat{\mathbf{u}}^n$ and $\hat{\mathbf{B}}^n$ and then solve VTBE to obtain the current solutions $\tilde{\mathbf{u}}^{n+1}$, $\Theta^{n+1}$ and $\tilde{\mathbf{B}}^{n+1}$.
	\item[Step 2:] Solve PPBE to obtain $P^{n+1}$ and $P^{n+1}_b$ based on $\tilde{\mathbf{u}}^{n+1}$ and $\tilde{\mathbf{B}}^{n+1}$.
	\item[Step 3:] Update the current solutions $\mathbf{u}^{n+1}$ and $\mathbf{B}^{n+1}$ according to equation \eqref{eqn:ubtilde}.
\end{description}
In the above time stepping scheme, the intermediate variables $\hat{\mathbf{u}}^n$ and $\hat{\mathbf{B}}^n$ are calculated from equations \eqref{eqn:ubhat} except the initial values $\hat{\mathbf{u}}^0$ and $\hat{\mathbf{B}}^0$ by first-order approximations $\hat{\mathbf{u}}^0 = \mathbf{u}^0$, $\hat{\mathbf{B}}^0 = \mathbf{B}^0$.

\subsection{Finite volume spatial discretization}
\label{spatialdiscretization}

\begin{figure}
	\centering
	\subfloat[]{\includegraphics[width=0.4\textwidth]{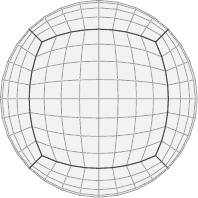}}
	\hfill
	\subfloat[]{\includegraphics[width=0.5\textwidth]{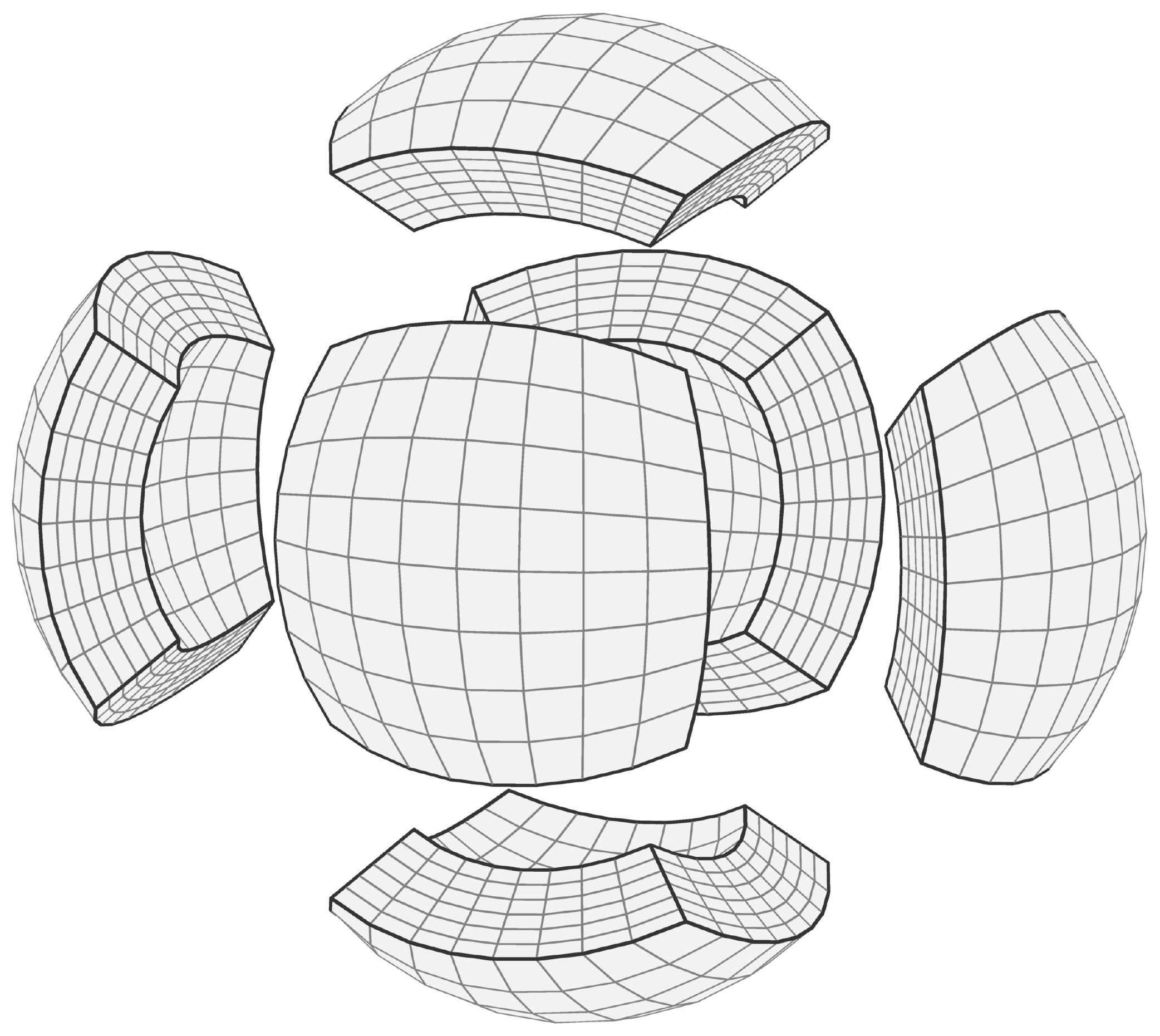}}
	\caption{A cubed-sphere grid based on the equiangular gnomonic projection. (a) $(\xi, \eta)$ grid on a spherical surface, (b) An open grid by shifting the six blocks outwards.}
	\label{fig:csmesh}
\end{figure}

As an alternative to the traditional latitude-longitude grid that suffers from disadvantages such as singularity and non-uniformity, the cubed-sphere grid \citep{Sadourny1972, Ronchi1996} obtained by a projection of the inscribed cube is becoming popular for problems defined on the spherical geometry.
Adopting the cubed-sphere grid based on the equiangular gnomonic projection \citep{Ronchi1996}, a spherical shell is divided into six identical blocks, of which each block is described by a local coordinate system $(\xi, \eta, r)$, $\xi, \eta \in [-\pi/4, \pi/4]$.
With each block being divided uniformly in the three coordinate directions, a quasi-uniform cubed-sphere grid covering the whole spherical shell can be obtained, as shown in Fig. \ref{fig:csmesh}.
In spite of the complexity caused by the non-orthogonality of $\xi$ and $\eta$, the resulting cubed-sphere grid is quite regular and thus can be adapted well to the algorithms of domain decomposition \citep{Toselli2005} and multigrid \citep{Saad2003}.

A collocated arrangement by which all the unknown variables $(\mathbf{u}, \Theta, \mathbf{B}, P, P_b)$ are located at the center of grid cells is employed in the spatial discretization.
For each block, the numbers of grid cells in $\xi$ and $\eta$ directions are set to be the same value $N_s$ and the cell number in $r$ direction is denoted by $N_r$.
The coordinates of the unknown point with the indices $(i, j, k), 0 \leq i, j \leq N_s-1, 0 \leq k \leq N_r-1$ in each block can be calculated by
\begin{equation}
	\xi_i = -\frac{\pi}{4} + (i + 0.5) h_s, \quad \eta_j = -\frac{\pi}{4} + (j + 0.5) h_s, \quad r_k = r_i + (k + 0.5) h_r,
\end{equation}
where $h_s = \pi/(2N_s)$, $h_r = (r_o - r_i)/N_r$ are the grid spacings in $\xi$ ($\eta$) and $r$ directions, respectively.
The total cell number of the cubed-sphere grid is denoted by $N = N_s \times N_s \times N_r \times 6$.

A finite volume scheme based on the cubed-sphere grid is applied to the spatial discretization of the linear operators in \eqref{eqn:linearoperator} and the nonlinear terms in \eqref{eqn:nonlinearoperator}.
Given a vector $\mathbf{v}$, its divergence at the center of the grid cell ($i, j, k$) is numerically approximated by the Gauss theorem
\begin{align}
	\label{eqn:divdiscretization}
	\left. \nabla \cdot \mathbf{v} \right|_{i,j,k} \approx {} & \frac{1}{V_{i,j,k}} \int_V \nabla \cdot \mathbf{v} \diff V = \frac{1}{V_{i,j,k}} \oint_S \mathbf{v} \cdot \diff \mathbf{S} 
	\notag\\ 
	\approx {} & \frac{1}{V_{i,j,k}} \left[ \left( v^1 \sqrt{g} \right)_{i + \frac 12, j, k} h_s h_r - \left( v^1 \sqrt{g} \right)_{i - \frac 12, j, k} h_s h_r \right.
	\notag\\
	& + \left( v^2 \sqrt{g} \right)_{i, j + \frac 12, k} h_s h_r - \left( v^2 \sqrt{g} \right)_{i, j - \frac 12, k} h_s h_r 
	\notag\\
	& + \left. \left( v^3 \sqrt{g} \right)_{i, j, k + \frac 12} h_s h_s - \left( v^3 \sqrt{g} \right)_{i, j, k - \frac 12} h_s h_s \right],
\end{align}
where $V_{i,j,k} \approx \sqrt{g}_{i,j,k} h_s h_s h_r$ refers to the volume of the grid cell ($i, j, k$), ($v^1, v^2, v^3$) are the contravariant components of $\mathbf{v}$ and $g$ is the determinant of the covariant components $g_{mn}$ of the metric tensor in the cubed-sphere grid
\begin{equation}
	\sqrt{g} = \sqrt{\mathrm{det}( g_{mn} )} = r^2 \sec^2\xi \sec^2\eta / \left( 1 + \tan^2\xi + \tan^2\eta \right)^{\frac{3}{2}}.
\end{equation}
The spatial differential operators in equations \eqref{eqn:linearoperator} and \eqref{eqn:nonlinearoperator} are transformed into the divergence forms and then discretized according to equation \eqref{eqn:divdiscretization}.

Most of the spatial terms in the governing equations \eqref{eqn:operatorvelocity}--\eqref{eqn:operatordivb} have been discussed in our previous work \citep{Yin2017}.
In this section, we focus on the new terms related to the magnetic field $\mathbf{B}$, including the Laplacian term $\nabla^2 \mathbf{B}$, divergence term $\nabla \cdot \mathbf{B}$ and nonlinear terms $\mathbf{B} \cdot \nabla \mathbf{B}$, $\mathbf{B} \cdot \nabla \mathbf{u}$ and $\mathbf{u} \cdot \nabla \mathbf{B}$.
The Laplacian term $\nabla^2 \mathbf{B}$ and divergence term $\nabla \cdot \mathbf{B}$ are discretized in the same way as $\nabla^2 \mathbf{u}$ and $\nabla \cdot \mathbf{u}$, respectively, while some additional effort is required for the three nonlinear terms.

To deal with the three nonlinear terms in a uniform way, we consider a generic form $\mathbf{a} \cdot \nabla \mathbf{b}$, where $\mathbf{a}$ and $\mathbf{b}$ are two arbitrary vectors conforming divergence-free condition $\nabla \cdot \mathbf{a} = \nabla \cdot \mathbf{b} = 0$.
The nonlinear term $\mathbf{a} \cdot \nabla \mathbf{b}$ can be rewritten in a conservative form and divided into two parts
\begin{equation}
	\mathbf{a} \cdot \nabla \mathbf{b} = \nabla \cdot \left( \mathbf{a} \mathbf{b} \right) = \left[ \nabla \cdot \left( \mathbf{a} b^k \right) + a^i b^j \Gamma^k_{ij} \right] \mathbf{g}_k, \quad i,j,k = 1,2,3,
\end{equation}
where $\Gamma^k_{ij} (i,j,k=1,2,3)$ are the Christoffel symbols whose expressions are
\begin{align}
	\left( \Gamma^1_{ij} \right) & = \begin{pmatrix}
		\dfrac{2\tan\xi\tan^2\eta}{1+\tan^2\xi+\tan^2\eta} & -\dfrac{\tan\eta\sec^2\eta}{1+\tan^2\xi+\tan^2\eta} & \dfrac1r \\[8pt]
		-\dfrac{\tan\eta\sec^2\eta}{1+\tan^2\xi+\tan^2\eta} & 0 & 0 \\[8pt]
		\dfrac1r & 0 & 0
	\end{pmatrix}, \\
	\left( \Gamma^2_{ij} \right) & = \begin{pmatrix} 
		0 & -\dfrac{\tan\xi\sec^2\xi}{1+\tan^2\xi+\tan^2\eta} & 0 \\[8pt]
		-\dfrac{\tan\xi\sec^2\xi}{1+\tan^2\xi+\tan^2\eta} & \dfrac{2\tan\eta\tan^2\xi}{1+\tan^2\xi+\tan^2\eta} & \dfrac1r \\[8pt]
		0 & \dfrac1r & 0
	\end{pmatrix}, \\
	\left( \Gamma^3_{ij} \right) & = \dfrac{r\sec^2\xi\sec^2\eta}{\left( 1+\tan^2\xi+\tan^2\eta \right)^2} \begin{pmatrix} 
		-\sec^2\xi & \tan\xi\tan\eta & 0 \\
		\tan\xi\tan\eta & -\sec^2\eta & 0 \\
		0 & 0 & 0
	\end{pmatrix},
\end{align}
and $\mathbf{g}_k (k=1,2,3)$ are the covariant base vectors in the cubed-sphere coordinate system.
The divergence term is discretized in a finite volume scheme
\begin{align}
	\left. \nabla \cdot \left( \mathbf{a} b^k \right) \right|_{i,j,k} \approx {} & \frac{1}{V_{i,j,k}} \int_V \nabla \cdot \left( \mathbf{a} b^k \right) \diff V = \frac{1}{V_{i,j,k}} \oint_S \left( \mathbf{a} b^k \right) \cdot \diff \mathbf{S} 
	\notag\\ 
	\approx {} & \frac{1}{V_{i,j,k}} \left[ \left( a^1 b^k \sqrt{g} \right)_{i + \frac 12, j, k} h_s h_r - \left( a^1 b^k \sqrt{g} \right)_{i - \frac 12, j, k} h_s h_r \right.
	\notag\\
	& + \left( a^2 b^k \sqrt{g} \right)_{i, j + \frac 12, k} h_s h_r - \left( a^2 b^k \sqrt{g} \right)_{i, j - \frac 12, k} h_s h_r 
	\notag\\
	& + \left. \left( a^3 b^k \sqrt{g} \right)_{i, j, k + \frac 12} h_s h_s - \left( a^3 b^k \sqrt{g} \right)_{i, j, k - \frac 12} h_s h_s \right].
	\label{eqn:nonlinearint}
\end{align}
The rest term can be expressed as
\begin{equation}
	\begin{aligned}
		a^i b^j \Gamma^1_{ij} & = a^1 b^1 \Gamma^1_{11} + a^1 b^2 \Gamma^1_{12} + a^1 b^3 \Gamma^1_{13} + a^2 b^1 \Gamma^1_{21} + a^3 b^1 \Gamma^1_{31}, \\
		a^i b^j \Gamma^2_{ij} & = a^1 b^2 \Gamma^2_{12} + a^2 b^1 \Gamma^2_{21} + a^2 b^2 \Gamma^2_{22} + a^2 b^3 \Gamma^2_{23} + a^3 b^2 \Gamma^2_{32}, \\
		a^i b^j \Gamma^3_{ij} & = a^1 b^1 \Gamma^3_{11} + a^1 b^2 \Gamma^3_{12} + a^2 b^1 \Gamma^3_{21} + a^2 b^2 \Gamma^3_{22},
	\end{aligned}
\end{equation}
and is treated as a source term.
We apply the above finite volume scheme to the three nonlinear terms $\mathbf{B} \cdot \nabla \mathbf{B}$, $\mathbf{B} \cdot \nabla \mathbf{u}$ and $\mathbf{u} \cdot \nabla \mathbf{B}$.

Some special attention should be paid to the boundary condition of the magnetic field.
Let $(B^1, B^2, B^3)$ denote the contravariant components of the magnetic field in the cubed-sphere grid.
According to the pseudo-vacuum boundary condition \eqref{eqn:pvc}, we can deduce that the tangential components of the magnetic field equal zero on the boundaries, i.e. $B^1 = B^2 = 0$.
The normal component $B^3$ can be constrained by the solenoidal condition $\nabla \cdot \mathbf{B} = 0$.
In the cubed-sphere grid, the solenoidal condition can be expressed as
\begin{equation}
	\nabla \cdot \mathbf{B} = \frac{\partial B^1}{\partial \xi} + \frac{\partial B^2}{\partial \eta} + \left( \Gamma^1_{11} + \Gamma^2_{12} \right) B^1 + \left( \Gamma^1_{12} + \Gamma^2_{22} \right) B^2 + \frac{1}{r^2} \frac{\partial}{\partial r} \left(r^2 B^3 \right) = 0.
\end{equation}
Due to $B^1 = B^2 = 0$, we can obtain
\begin{equation}
	\frac{\partial}{\partial r} \left(r^2 B^3 \right) = 0, 
\end{equation}
following which the normal component of magnetic field $B^3$ on the boundaries is calculated.

\subsection{Parallel solution and multilevel preconditioner}
At each time step, there are two linear algebraic equations, i.e. VTBE and PPBE, to be considered.
The Krylov subspace iterative method combined with the preconditioning technology is employed to solve these linear systems in this paper.
With preconditioning, a linear system, e.g. $Ax=b$, is replaced with a right preconditioned system
\begin{equation}
	\label{eqn:newlinear}
	A' x' = b,
\end{equation}
where $A' = A M^{-1}, x' = M x$. Here the matrix $M$ is generally called preconditioner.
For any time step, $x'$ is initialized as $x'_0 = M x_0$ where $x_0$ is usually set to be the solution of previous time step.
Then the new preconditioned linear system \eqref{eqn:newlinear} is solved by a restarted generalized minimum residual (GMRES) algorithm until the residual satisfies
\begin{equation}
	\left\| A' x' - b \right\| \leq \max \{\varepsilon_a, \varepsilon_r \left\| A' x'_0 - b \right\| \},
\end{equation}
where $\varepsilon_a$, $\varepsilon_r$ are the absolute and relative convergence tolerances, respectively.
And finally the present time step solution $x$ can be obtained by $x = M^{-1} x'$.

When solving the linear system \eqref{eqn:newlinear} by the Krylov subspace iterative method, the convergence rate strongly depends on the condition number of the coefficient matrix $A'= A M^{-1}$ \citep{Demmel1997}.
If $A'$ is well conditioned, that is, its condition number is sufficiently small, the iteration number of the Krylov subspace method can be dramatically reduced.
This can be achieved by choosing an appropriate preconditioner $M$.
A good choice of the preconditioner should also help improve the scalability of parallel computations on large-scale supercomputers.
In other words, with the aid of a scalable preconditioner, the iteration number should remain a steady level as the number of processor cores increases.
It is often problem-dependent to construct an efficient and scalable preconditioner.
In present study, we design a parallel multi-level restricted additive Schwarz preconditioner based on domain decomposition and multigrid method.

\begin{figure}
	\centering
	\subfloat[]{\includegraphics[width=0.45\textwidth]{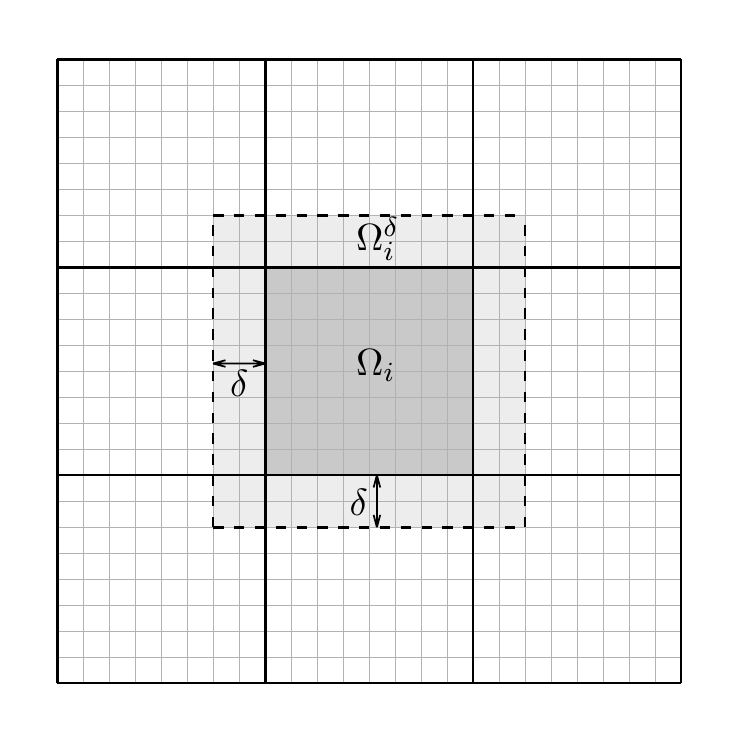} \label{fig:subdomain}}
	\subfloat[]{\includegraphics[width=0.45\textwidth]{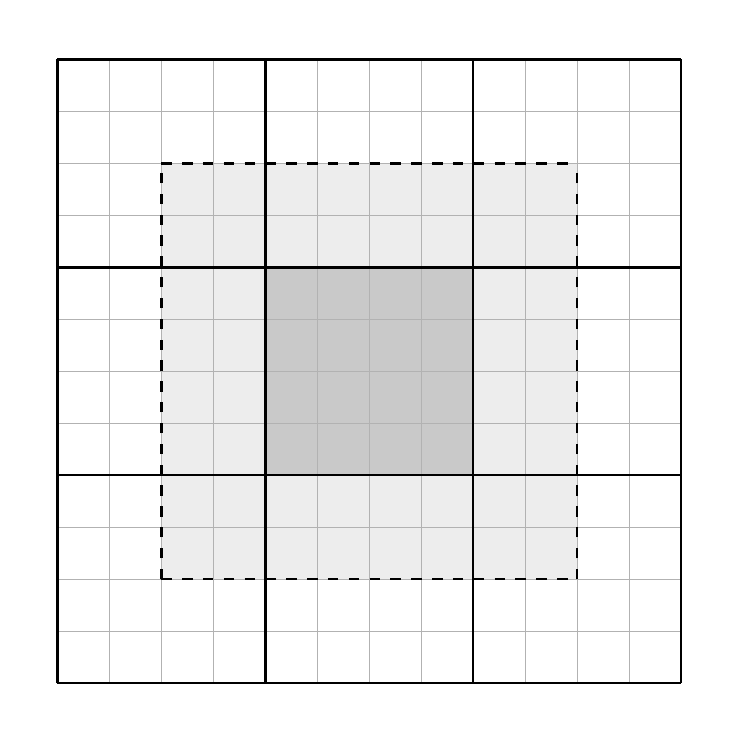}}
	\caption{A two-dimensional illustration of domain decomposition and multigrid. (a) fine mesh, (b) coarse mesh of which the cell number in each direction is reduced by $1/2$.}
	\label{fig:decomposition}
\end{figure}

The cubed-sphere grid is divided into six identical blocks and each block is decomposed into $p=p_1 p_2 p_3$ non-overlapping subdomains in a structured manner, where $p_1$, $p_2$, $p_3$ are numbers of subdivisions corresponding to three coordinate directions respectively.
Each subdomain is assigned to one processor core and the number of processor cores corresponding to each block is $p$.
Thus the total number of processor cores is $6p$ as well as the total number of subdomains.
For each non-overlapping subdomain $\Omega_i$, $i = 1, 2, \ldots, 6p$, we can obtain a corresponding larger overlapping subdomain $\Omega^\delta_i$ by extending $\Omega_i$ with $\delta$ layers of mesh cells, as shown in Fig. \ref{fig:subdomain}.
The subdomains containing one or more block interfaces are extended to the adjacent mesh cells of the neighbouring block(s).
The extending parts of overlapping subdomains lead to data exchanges, i.e. communications between corresponding processor cores.
To achieve good scalability, the influence of communication time should be reduced as much as possible.

Let $N$ denote the total number of mesh cells and $d$ be the number of degrees of freedom per point.
Moreover, the number of mesh cells in overlapping subdomain $\Omega_i^\delta$ is denoted by $N_i^\delta$.
Then we can define a one-level restricted additive Schwarz (RAS) \citep{Cai1999} preconditioner as
\begin{equation}
	M^{-1}_{\mathrm{one}} = \sum^{6p}_{i=1} \left( R^0_i \right)^\mathrm{T} \left( A^\delta_i \right)^{-1} R^\delta_i.
	\label{eqn:onelevel}
\end{equation}
The restriction operator $R^\delta_i$ is an $N^\delta_i \times N$ block matrix and its multiplication by a $N \times 1$ block vector defined on the entire domain results in a smaller $N^\delta_i \times 1$ block vector defined on the subdomain $\Omega^\delta_i$ by dropping the components corresponding to the mesh cells outside $\Omega^\delta_i$.
The element of the restriction matrix $\left( R^\delta_i \right)_{q_1, q_2}$, which is a $d \times d$ block submatrix, is an identity block if the integer indices $1\leq q_1 \leq N^\delta_i$ and $1 \leq q_2 \leq N$ belong to a cell in the overlapping subdomain $\Omega^\delta_i$, or a block of zeros otherwise.
As a special case, $R^0_i$ is also an $N^\delta_i \times N$ block matrix that is similarly defined, but is a restriction to the non-overlapping subdomain $\Omega_i$.
The matrix $A^\delta_i$ is the restriction of the coefficient matrix $A$ to the overlapping subdomain $\Omega^\delta_i$ with size $N^\delta_i \times N^\delta_i$ and is defined as $A^\delta_i = R^\delta_i A \left( R^\delta_i \right)^\mathrm{T}$.
The matrix-vector multiplication with $\left( A^\delta_i \right)^{-1}$ refers to solving a local linear system in subdomain $\Omega^\delta_i$ and can be computed exactly by using a sparse LU factorization.
Since LU factorization is often expensive and to form an exact preconditioner is generally not necessary, the matrix-vector multiplication is usually obtained approximately by a less expensive incomplete LU (ILU) factorization.

In our previous work \citep{Yin2017}, it is found that the one-level RAS preconditioner can achieve very good parallel performance for the solution of the velocity-related equation but scales poorly for the pressure-related equation.
To improve the scalability of the one-level RAS preconditioner, we employ a multi-level RAS method based on hybrid preconditioning \citep{Mandel1994} and multigrid technique.
By combining the one-level RAS preconditioner $B_f$ with a coarse level preconditioner $B_c$ defined on a coarser mesh in a multiplicative manner, we obtain a hybrid preconditioner
\begin{equation}
	M^{-1}_{\mathrm{two}} = \mathrm{hybrid} \left( B_c, B_f \right) = B_c + B_f - B_f A_f B_c,
	\label{eqn:twolevel}
\end{equation}
where $B_c = \mathcal{I}^f_c A^{-1}_c \mathcal{I}^c_f$ and $A_f$, $A_c$ denote the coefficient matrices on the fine and coarse meshes, respectively.
Here, $\mathcal{I}_f^c$ is a restriction operator mapping from a vector defined on the fine mesh to a coarse mesh vector.
Similarly, $\mathcal{I}_c^f$ is a prolongation operator from the coarse mesh to the fine mesh.
More precisely speaking, to calculate the multiplication of the hybrid two-level preconditioner and a vector $x$, $y=M^{-1}_{\mathrm{two}} x$, we first apply a coarse mesh preconditioning
\begin{equation}
	w = \left( \mathcal{I}_c^f A^{-1}_c \mathcal{I}_f^c \right) x, \label{eqn:coarsesolve}
\end{equation}
and then correct the coarse solution by adding the fine mesh solution to obtain the final result
\begin{equation}
	y = w + M^{-1}_{\mathrm{one}} \left( x - A_f w \right). \label{eqn:finesolve}
\end{equation}

For each application of the two-level preconditioner \eqref{eqn:twolevel}, a smaller linear system with the coefficient matrix $A_c$ on the coarse mesh needs to be dealt with during the coarse mesh preconditioning.
This coarse level linear system is solved by using preconditioned GMRES with a relative tolerance $\eta_c$.
The coarse level preconditioner can be either one-level \eqref{eqn:onelevel} or two-level \eqref{eqn:twolevel}.
When a two-level preconditioner is adopted on the coarse mesh as well, another coarser mesh is required to form this preconditioner.
Repeating the application of the two-level RAS preconditioner \eqref{eqn:twolevel} in multiple mesh levels can result in a multilevel hybrid RAS method.

The best choices for some of the options in the multilevel RAS preconditioner are often problem-dependent \citep{Yang2011a}.
One important option is the number of mesh levels, whose choice strongly depends on a specific circumstance.
Since additional computational costs can be introduced by the coarse meshes, excessive mesh levels may lead to the degradation of computational efficiency.
Furthermore, the choice of the number of mesh levels has a close relationship to the problem size.
If too many mesh levels are applied when the problem size is not large enough, the computational load of each processor core may be too small.
At this situation, the influence of communication time may become remarkable and the scalability may become worse.
In the present study, we choose a two-level version with a coarse-to-fine mesh ratio 1:2 in each direction (see Fig. \ref{fig:decomposition}), by which an optimal efficiency is achieved in the considered spatial resolutions.
If a larger resolution is required, three or more levels may be taken into consideration to achieve better performance.
On the fine level mesh denoted by $N$, the preconditioner is 
\begin{equation}
	M^{-1}_N = \mathrm{hybrid} \left( \mathcal{I}_{N/2}^N A^{-1}_{N/2} \mathcal{I}^{N/2}_N, \sum^{6p}_{i=1} \left( (R_N)^0_i \right)^\mathrm{T} \left( (A_N)^\delta_i \right)^{-1} (R_N)^\delta_i \right),
	\label{eqn:finepreconditioner}
\end{equation}
where $N/2$ refers to the coarse level mesh.
A linear average restriction operator $\mathcal{I}^{N/2}_N$ and a piecewise constant interpolation operator $\mathcal{I}^N_{N/2}$ are employed due to their simplicities.
The linear system about $A_{N/2}$ on the coarse level mesh is solved by an inner GMRES, preconditioned by a one-level RAS approach on the corresponding coarse level
\begin{equation}
	M^{-1}_{N/2} = \sum^{6p}_{i=1} \left( (R_{N/2})^0_i \right)^\mathrm{T} \left( (A_{N/2})^\delta_i \right)^{-1} (R_{N/2})^\delta_i.
	\label{eqn:coarsepreconditioner}
\end{equation}
The choice of the subdomain solver at each level has a strong influence on the overall performance of the preconditioner.
A large number of numerical experiments are often necessary to find out the proper selection.
According to our tests, the ILU factorization with no fill-in, ILU(0), is chosen as the subdomain solver for both the fine level $\left( (A_{N})^\delta_i \right)^{-1}$ in equation \eqref{eqn:finepreconditioner} and the coarse level $\left( (A_{N/2})^\delta_i \right)^{-1}$ in equation \eqref{eqn:coarsepreconditioner}.

\section{Numerical results}
\label{numericalresults}

We build the parallel simulation code based on the Portable, Extensible Toolkit for Scientific Computation (PETSc) library \citep{Balay2013} and carry out the numerical experiments on the Sunway TaihuLight supercomputer \citep{Fu2016} which took the top place of the Top-500 list \citep{Top500} as of June 2016.
The two resulting sparse linear algebraic equations, i.e. VTBE and PPBE, are solved by GMRES algorithm with the restarting parameter 30.
The absolute and relative tolerance of GMRES are respectively set to be $10^{-10}$, $10^{-8}$ for VTBE and $10^{-8}$, $10^{-6}$ for PPBE.

\subsection{Convergence test}
\begin{figure}
	\centering
	\subfloat[]{\includegraphics[width=0.5\textwidth]{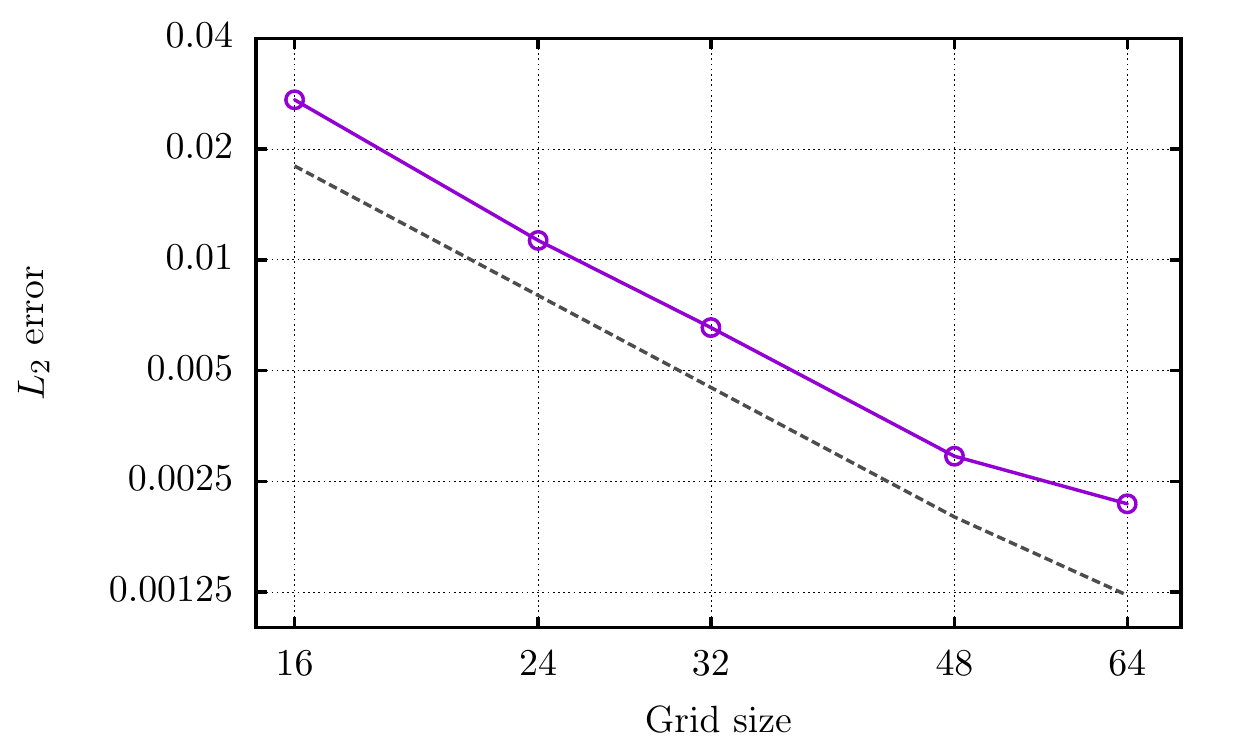} \label{fig:space_convergence}}
	\subfloat[]{\includegraphics[width=0.5\textwidth]{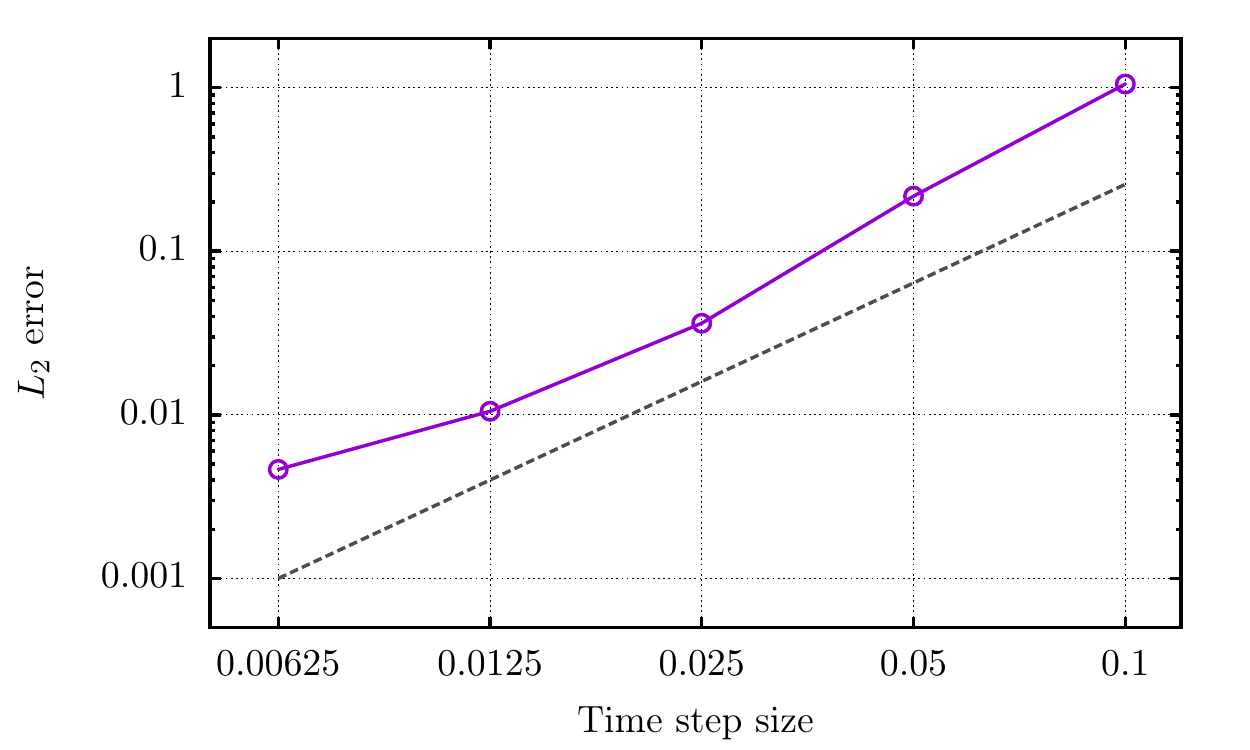} \label{fig:time_convergence}}
	\caption{Convergence analysis in terms of the spatial accuracy and the temporal accuracy. (a) The $L_2$ error of $(\mathbf{u}, \Theta, \mathbf{B})$ with respect to different grid sizes. The reference solution is obtained on a $128\times128\times128\times6$ mesh. The horizontal axis shows the grid size in one direction. The time step size is fixed to $1\times10^{-5}$ for all grid sizes. 
	(b) The $L_2$ error of $(\mathbf{u}, \Theta, \mathbf{B})$ with respect to different time step sizes. The reference solution is obtained with $\Delta t = 3.125\times10^{-3}$. The spatial resolution is $64\times64\times64\times6$ for all time step sizes.
	The two gray dashed lines indicate the ideal second-order convergence.}
	\label{fig:accuracy_convergence}
\end{figure}

To validate the accuracy of the spatial discretization and temporal integration scheme, a convergence analysis is performed for the dynamo problem with the parameters $Pr = 1, Ra = 100, E = 0.1, Pm = 1000$ and the initial condition \eqref{eqn:initial}.
We define the $L_2$ error \citep{Leveque2002book} of a solution $v$ as 
\begin{equation}
	L_2(v) = \sqrt{\sum_{l=1}^{6}\sum_{i,j,k} \sum_{c} (v_{l,i,j,k,c} - v_{l,i,j,k,c}^{ref})^2 V_{i,j,k} },
\end{equation}
where $l$ is the block index, $c$ is the component index in each mesh cell, $(i,j,k)$ are the grid indices, $v^{ref}$ is the reference solution and $V_{i,j,k} \approx \sqrt{g}_{i,j,k} h_s h_s h_r$ refers to the volume of the grid cell ($i, j, k$).
In order to quantify the error of the spatial discretization and cancel out the temporal error, we fix the time step size to $\Delta t = 1 \times 10^{-5}$ and adopt the solution at $t=0.001$ on $128\times128\times128\times6$ mesh as the reference solution.
The $L_2$ errors of $(\mathbf{u}, \Theta, \mathbf{B})$ with respect to different grid sizes are provided in Fig. \ref{fig:space_convergence}, in which we also plot the ideal second-order convergence line.
It can be seen that the spatial discretization is second-order accurate.
In terms of the temporal convergence, we fix the spatial resolution to be $64\times64\times64\times6$ and use the solution at $t = 1$ with the time step size $\Delta t = 3.125\times10^{-3}$ as the reference solution.
The $L_2$ errors of $(\mathbf{u}, \Theta, \mathbf{B})$ with respect to different time step sizes are shown in Fig. \ref{fig:time_convergence}, where the ideal second-order convergence line is also provided.
We can observe from the figure that second-order accuracy is achieved with the employed the temporal integration scheme.

\subsection{Benchmark cases}
Following the well-known benchmark study of \citet{Christensen2001} where the insulating boundary condition is considered, a shell dynamo benchmarking exercise with pseudo-vacuum boundary conditions was carried out by \citet{Jackson2014} for the first time.
Under the parameter regime in their work,
\begin{equation}
	E = 10^{-3}, \quad Pr = 1, \quad Ra = 100, \quad Pm = 5,
\end{equation}
at least five magnetic diffusion times are required to reach the quasi-steady state from the suggested initial condition
\begin{equation}
	\label{eqn:initial}
	\begin{aligned}
		\mathbf{u}^0 & = 0, \\
		\Theta^0 & = \frac{21}{\sqrt{17920\pi}} \left( 1 - 3 x^2 + 3 x^4 - x^6 \right) \sin^4 (\theta) \cos (4\phi), \quad x = 2 r - r_i - r_o, \\
		B^0_r & = \frac 58 \frac{9 r^3 - 4 [ 4 + 3 (r_i + r_o) ] r^2 + [ 4 r_o + r_i (4 + 3 r_o) ] 6 r - 48 r_i r_o}{r} \cos (\theta), \\
		B^0_\theta & = - \frac{15}{4} \frac{(r-r_i)(r-r_o)(3r-4)}{r} \sin (\theta), \\
		B^0_\phi & = \frac{15}{8} \sin [\pi (r-r_i)] \sin (2\theta).
	\end{aligned}
\end{equation}
In the same spirit, \citet{Vantieghem2016} suggests a new benchmark case with the non-dimensional control parameter
\begin{equation}
	E = 10^{-3}, \quad Pr = 1, \quad Ra = 100, \quad Pm = 8,
\end{equation}
and the same initial condition for the dynamo validations with pseudo-vacuum boundary conditions.
According to their numerical results, a quasi-steady solution can be reached within less than one magnetic diffusion time.
In this section, we follow both these benchmark cases to validate the correctness of our finite volume code.
For simplicity, we refer to the benchmark case proposed by \citet{Jackson2014} as case P5 and the case suggested by \citet{Vantieghem2016} as case P8.

The values of the magnetic energy, kinetic energy and some other quantities at the final quasi-steady state are compared with the reference solutions for the benchmark case P5 and P8.
To compare with the benchmark results, these quantities should be calculated in a consistent dimension and we transform the present solutions into the dimensions that are consistent with \citet{Jackson2014} and \citet{Vantieghem2016}.
The relevant transformation formulae can be easily obtained according to the conversion table in \citep[][Table 1]{Jackson2014}.
Thus, for the purpose of consistency, the kinetic energy $E_{\mathrm{kin}}$ and magnetic energy $E_{\mathrm{mag}}$ are defined as follows
\begin{align}
	&E_{\mathrm{kin}} = \frac{Pm^2}{2} \int \mathbf{u}^2 \diff V, \\
	&E_{\mathrm{mag}} = \frac{Pm}{2 E} \int \mathbf{B}^2 \diff V.
\end{align}
And the velocity, temperature and magnetic field are calculated by transforming from the quantities in present dimension
\begin{align}
	&T' = T, \\
	&\mathbf{u}' = \mathbf{u} Pm, \\
	&\mathbf{B}' = \mathbf{B} / \sqrt{2}.
\end{align}

To conduct the numerical experiments for the case P5 and P8, four different levels of spatial resolution, i.e. G48, G64, G80 and G96, are considered here.
The mesh sizes, the overall resolutions and the corresponding time step sizes for these four employed grid levels are summarized in Table \ref{tab:gridlevels}, where the overall resolution $R$ is defined as the third root of the number of degrees of freedom for each scalar variable $R = N_{\mathrm{grid}}^{1/3}$.
It is worth mentioning that the explicit treatment of the nonlinear terms in the temporal integration scheme leads to a limitation of the time step size due to numerical stability condition.
To test the stability constraint, we define the Courant-Friedrichs-Lewy (CFL) number on the Alfv\'en waves \citep{Goedbloed2010}
\begin{equation}
	CFL_A = \max\left(\left|\frac{v_A^1}{h_s}\right| + \left|\frac{v_A^2}{h_s}\right| + \left|\frac{v_A^3}{h_r}\right|\right) \Delta t,
\end{equation}
where $\mathbf{v}_A = \mathbf{B} / \sqrt{E Pm}$ is the dimensionless Alfv\'en velocity.
In Table \ref{tab:gridlevels}, we also display the $CFL_A$ evaluated at the initial state on the four grid levels, within which no numerical instability is observed.
The time step sizes are carefully chosen to obey the stability constraint for the case P5 and P8 on such grids, as we find that the increase of $CFL_A$ by around 10\% may result in numerical instability.

\begin{table}
	\caption{Summary of the four employed grid levels. The column of total cells provides the total number of grid cells. $R$ is the overall resolution defined as the third root of the number of degrees of freedom for each scalar variable $R = N_{\mathrm{grid}}^{1/3}$.}
	\label{tab:gridlevels}
	\centering
	\begin{tabular}{lllllllll}
		\toprule
		Grid levels &  $N_s$ & $N_r$ & Total cells & $R$ & $\Delta t$ (P5)       & $CFL_A$ (P5) & $\Delta t$ (P8)      & $CFL_A$ (P8)     \\
		\midrule
		G48         &  48    & 72    & 995328      & 100 & $6.4 \times 10^{-5}$  & 1.13         & $8.0 \times 10^{-5}$  & 1.12 \\
		G64         &  64    & 96    & 2359296     & 133 & $4.0 \times 10^{-5}$  & 0.95         & $5.0 \times 10^{-5}$  & 0.94 \\
		G80         &  80    & 120   & 4608000     & 166 & $4.0 \times 10^{-5}$  & 1.19         & $5.0 \times 10^{-5}$  & 1.18 \\
		G96         &  96    & 144   & 7962624     & 200 & $3.2 \times 10^{-5}$  & 1.15         & $4.0 \times 10^{-5}$  & 1.13 \\
		\bottomrule
	\end{tabular}
\end{table}

\begin{figure}
	\centering
	\includegraphics[width=0.9\textwidth]{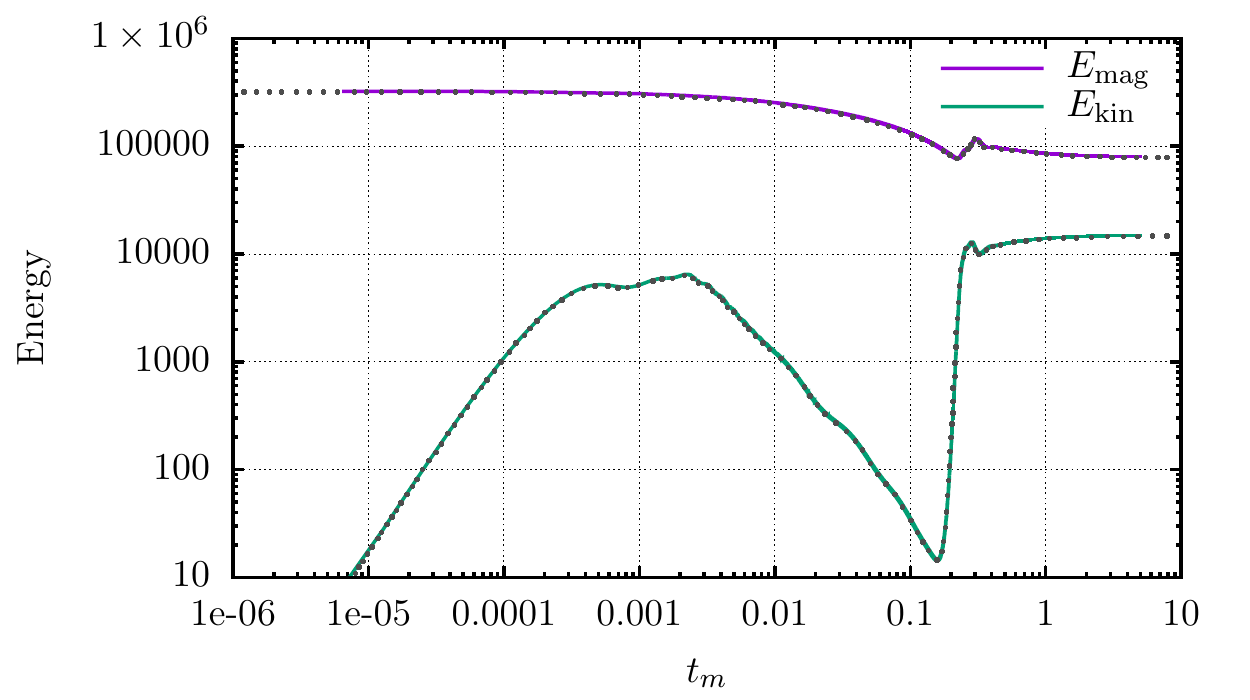}
	\caption{Time evolution of the magnetic energy $E_{\mathrm{mag}}$ and the kinematic energy $E_{\mathrm{kin}}$ on grid G96 for the benchmark case P5. The magnetic time $t_m$ is measured in units of magnetic diffusion time $t_m = t/Pm$. The black dots indicate the reference data taken from \citet{Jackson2014}.}
	\label{fig:EmagPm5}
\end{figure}

\begin{figure}
	\centering
	\includegraphics[width=0.9\textwidth]{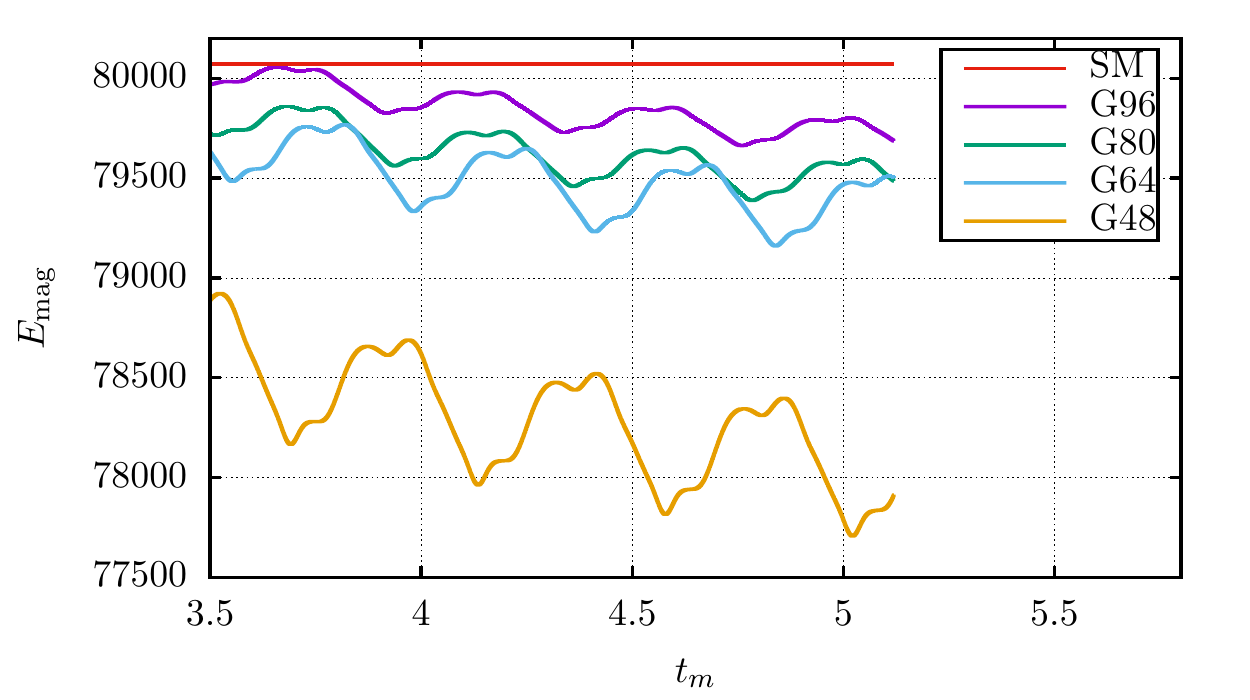}
	\caption{Slight oscillations of $E_{\mathrm{mag}}$ introduced by the drift of quasi-steady solution with respect to the grid on the four grid levels. SM denotes the recommended value of $E_{\mathrm{mag}}$ at $t_m = 5.12$ \citep{Jackson2014}.}
	\label{fig:EmagPm5meshOsci}
\end{figure}

Firstly, the benchmark case P5 is run by our finite volume code in the four different spatial resolutions listed in Table \ref{tab:gridlevels} until $t=25.6$, when the magnetic time measured in units of magnetic diffusion time is $t_m = t/Pm = 5.12$.
The time step sizes corresponding to the employed grid levels can be found in Table \ref{tab:gridlevels} as well.
The time evolution of the magnetic energy $E_{\mathrm{mag}}$ and the kinetic energy $E_{\mathrm{kin}}$ on grid G96 is displayed in Fig. \ref{fig:EmagPm5}, which shows good agreement with the result in \cite{Jackson2014}.
As discussed in \citet{Harder2005} and \citet{Jackson2014}, the drift of the quasi-steady solution with respect to the grid may introduce a slight temporal oscillation into the numerical solution.
The periodicity of this grid-drift oscillation, depending on the spatial periodic property of grid, is $90^\circ$ for the cubed-sphere grid.
In Fig. \ref{fig:EmagPm5meshOsci}, we display the oscillations of $E_{\mathrm{mag}}$ on the four grid levels by limiting $t_m$ to a small local time range near $t_m = 5.12$.
From Fig. \ref{fig:EmagPm5meshOsci}, it is seen that the amplitude of the grid-drift oscillation decays as the spatial resolution increases.
Note that this grid-drift oscillation only occurs for the steadily drifting solution, which is specifically designed for the benchmark purposes, and should vanish for the real dynamo simulations.
In addition, the solution of the case P5 is close to the onset of stable dynamo action and a slow decreasing observed from Fig. \ref{fig:EmagPm5meshOsci} indicates that the stable solution is gradually settling in near $t_m = 5.12$.
Since the decreasing is very slow (approximately 0.07\%), the numerical result at $t_m = 5.12$ is quite close to the stable solution.

\begin{figure}
	\centering
	\subfloat[]{\includegraphics[width=0.5\textwidth]{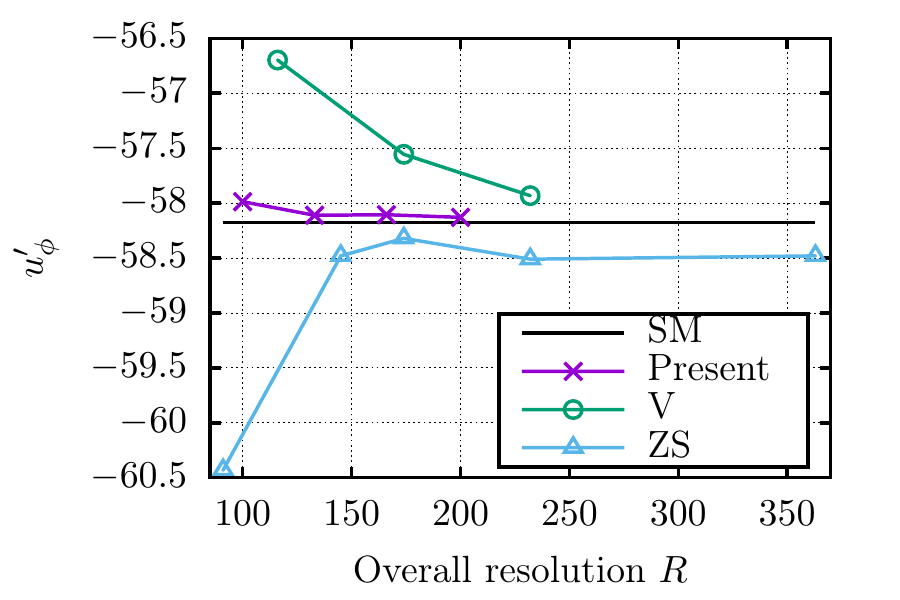}}
	\subfloat[]{\includegraphics[width=0.5\textwidth]{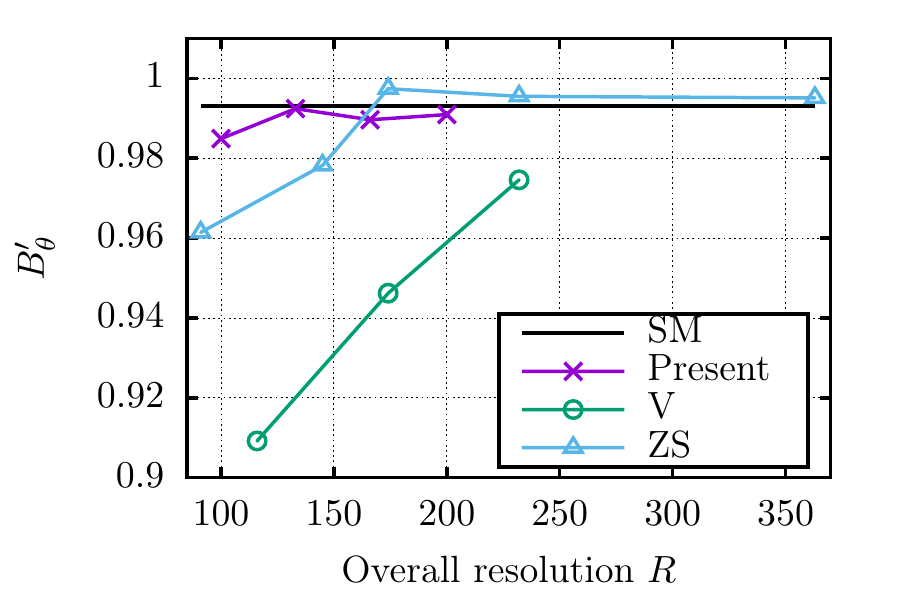}}
	\\
	\subfloat[]{\includegraphics[width=0.5\textwidth]{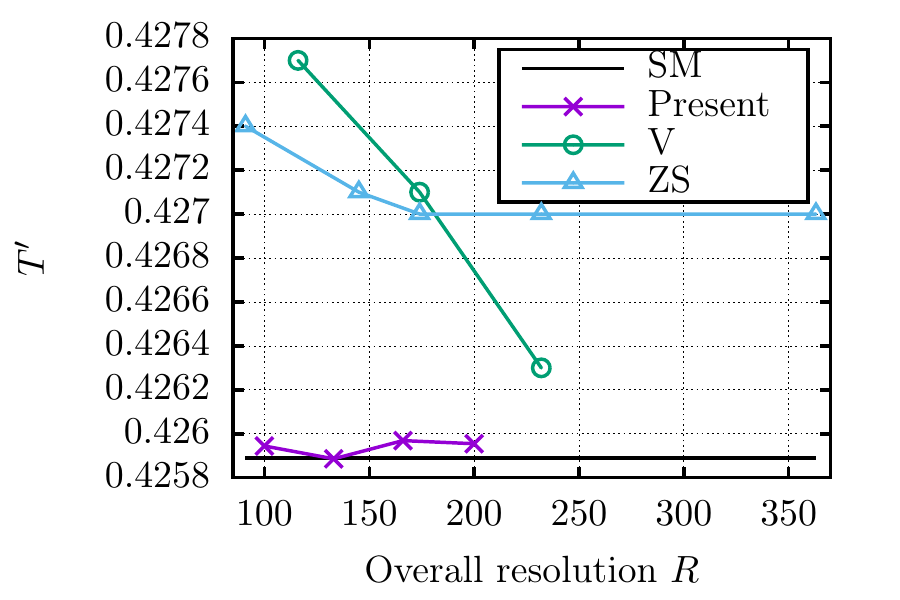}}
	\subfloat[]{\includegraphics[width=0.5\textwidth]{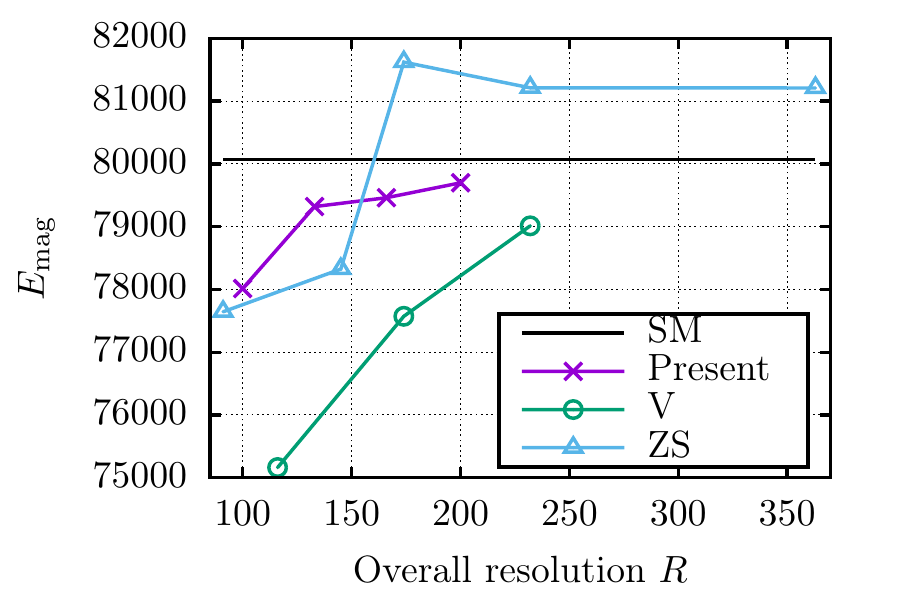}}
	\\
	\subfloat[]{\includegraphics[width=0.5\textwidth]{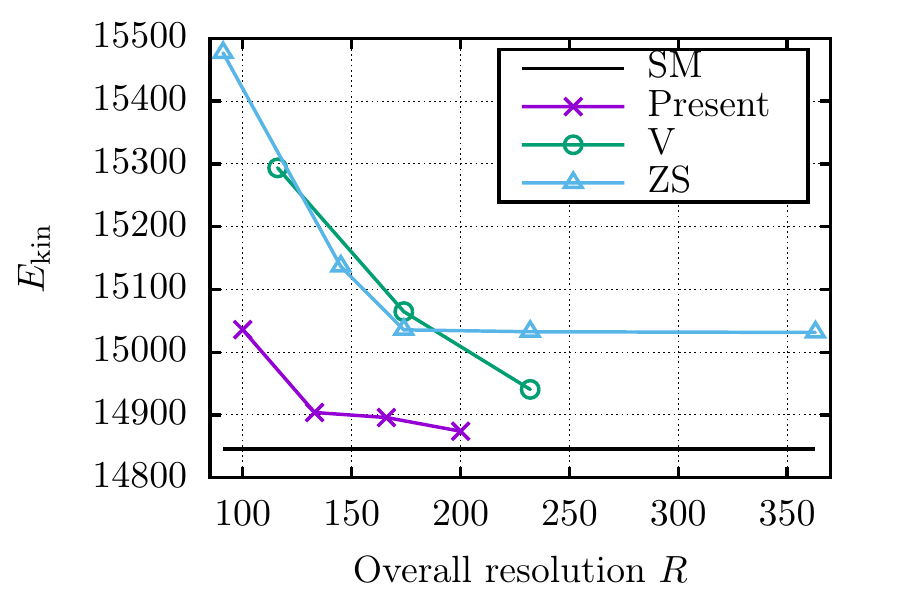}}
	\subfloat[]{\includegraphics[width=0.5\textwidth]{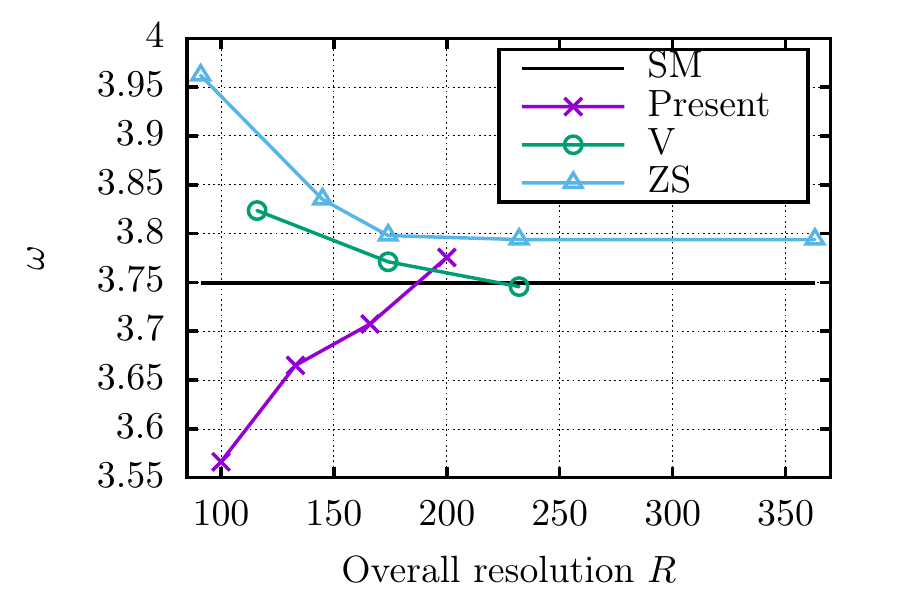}}
	\caption{Convergences as a function of the overall resolution $R$ for the case P5. (a) $u'_\phi$, (b) $B'_\theta$, (c) $T'$, (d) $E_{\mathrm{mag}}$, (e) $E_{\mathrm{kin}}$, (f) $\omega$. The present result and the recommended solution by spectral methods are denoted by Present and SM, respectively. V and ZS respectively refer to the convergence results of the finite volume code V and the finite element code ZS reported in \citet{Jackson2014}. }
	\label{fig:convergencePm5}
\end{figure}

\begin{table}
	\caption{Comparison with the benchmark results for the case P5. G48, G64, G80 and G96 are our numerical results in four different spatial resolutions. SM denotes the recommended benchmark solution \citep{Jackson2014} obtained by the spectral methods. V232 and ZS363 are respectively the largest resolution results obtained by the finite volume code V at the overall resolution $R=232$ and the finite element code ZS at $R=363$ reported in \citet{Jackson2014}. SFEMaNS refers to the finite element result reported in \citet{Matsui2016}. The values in parentheses are relative errors compared with the SM results.}
	\label{tab:comparisonPm5}
	\centering
	\footnotesize
	\begin{tabular}{llllllll}
		\toprule
		Results  & $R$ & $E_{\mathrm{mag}}$ & $E_{\mathrm{kin}}$ & $T'$ & $u'_\phi$ & $B'_\theta$ & $\omega$ \\
		\midrule
	 	G48    & 100 &  78015(2.57\%) & 15036(1.28\%) & 0.425945(0.013\%) & $-57.987(0.33\%)$ & 0.9849(0.81\%) & 3.5664(4.88\%) \\
		G64    & 133 &  79320(0.94\%) & 14904(0.39\%) & 0.425886(0.001\%) & $-58.110(0.12\%)$ & 0.9924(0.06\%) & 3.6651(2.25\%) \\
		G80    & 166 &  79461(0.76\%) & 14896(0.34\%) & 0.425969(0.019\%) & $-58.105(0.13\%)$ & 0.9896(0.35\%) & 3.7075(1.12\%) \\
		G96    & 200 &  79700(0.46\%) & 14874(0.19\%) & 0.425955(0.015\%) & $-58.130(0.08\%)$ & 0.9909(0.21\%) & 3.7756(0.70\%) \\
		\\
		V232   & 232 & 79012(1.32\%) & 14941(0.64\%) & 0.42630(0.096\%) & $-57.932(0.42\%)$ & 0.9746(1.85\%) & 3.7457(0.10\%) \\
		ZS363  & 363 & 81210(1.42\%) & 15032(1.25\%) & 0.42700(0.261\%) & $-58.480(0.52\%)$ & 0.9951(0.21\%) & 3.7940(1.19\%) \\
		SFEMaNS&     & 80578(0.63\%) & 14797(0.33\%) & 0.42553(0.085\%) & $-58.280(0.17\%)$ & 1.0015(0.86\%) &  \\
		SM     &     & 80071 & 14846 & 0.42589 &  $-58.179$ & 0.9930 & 3.7495 \\
		\bottomrule
	\end{tabular}
\end{table}
From the quasi-steady solution, we calculate the final global data of the kinetic energy $E_{\mathrm{kin}}$, magnetic energy $E_{\mathrm{mag}}$ and drift frequency $\omega$, and the local data of $T'$, $u'_\phi$ and $B'_\theta$ at a reference point in the equatorial plane at mid-depth where $u_r=0$ and $(\partial u_r / \partial \phi) > 0$. 
To eliminate the influence of the grid-drift oscillation, these values are averaged over the period when the location of the local reference point changes by $90^\circ$, as was done in \citet{Harder2005} and \citet{Jackson2014}.
The average results are reported in Table \ref{tab:comparisonPm5}, where comparisons with the recommended benchmark solution obtained by using spectral methods and three other results with local methods are also provided.
V232 and ZS363 are respectively the largest resolution results obtained by the finite volume code V at the overall resolution $R=232$ and the finite element code ZS at $R=363$ reported in \citet{Jackson2014}.
And SFEMaNS refers to the finite element result reported in \citet{Matsui2016}.
The values in parentheses denote relative errors compared with the recommended benchmark solution obtained by spectral methods.
From Table \ref{tab:comparisonPm5}, it is seen that the discrepancy is less than 1\% for all quantities on grid G64, G80 and G96 except the drift frequency $\omega$.
On the coarsest grid G48, some of the relative errors are slightly large but less than 5\% and reduce considerably as the grid gets finer.
As for the error of $\omega$, good convergence rate with respect to the spatial resolution can be observed and the discrepancy drops below 1\% on grid G96.
The drift frequency $\omega$ is usually the most difficult to determine precisely in local methods, since it is usually obtained by interpolating from the discrete solution data whereby additional error source could be introduced.
Noticing that the overall resolution of the grid G64, G80 and G96 are respectively 133, 166 and 200, our finite volume code produces highly accurate solutions, which are comparable to and even better than the existing local results, for the benchmark case P5.
The convergences of these quantities as a function of the overall resolution $R$ are plotted in Fig. \ref{fig:convergencePm5}, in which we also provide the convergence results of the finite volume code V and the finite element code ZS reported in \citet{Jackson2014} for the purpose of comparison.

The benchmark case P8 is then considered to further validate the proposed methods and the implemented finite volume code.
An attractive advantage of this benchmark is that a quasi-steady solution can be reached within one magnetic diffusion time, which allows a much quicker validation in contrast to the benchmark case P5.
It was found by \citet{Sheyko2014} that two different types of dynamo solutions can be obtained when changing the initial magnetic field for this benchmark  problem.
For initial values of the magnetic energy between 407101 and 623428, such as the suggested one \eqref{eqn:initial}, one can obtain a quasi-steady solution expressed in the form $(\mathbf{u}, \mathbf{B}, \Theta) = f (r, \theta, \phi - \omega t)$.
For the initial magnetic energy outside this range, an oscillating dynamo solution can be found \citep{Vantieghem2016}.

\begin{figure}
	\centering
	\includegraphics[width=0.9\textwidth]{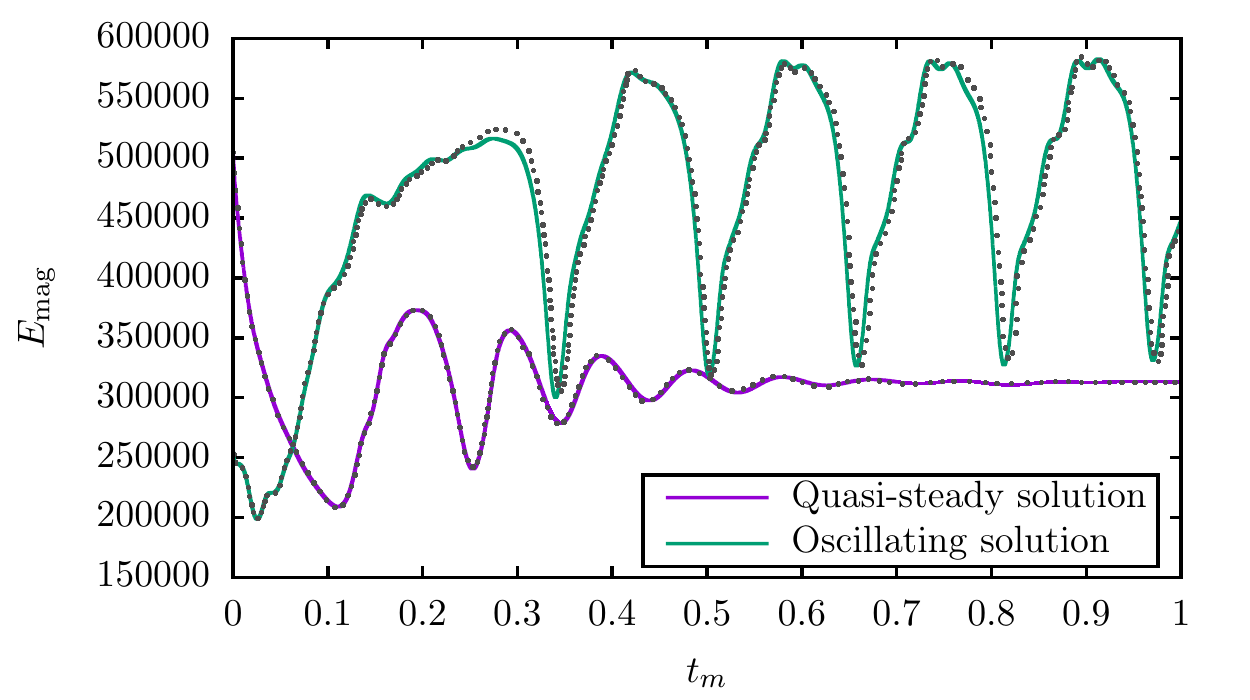}
	\caption{Time evolution of the magnetic energy $E_{\mathrm{mag}}$ on the grid G64 for the benchmark case P8 with two different initial magnetic field intensities. The magnetic time $t_m$ is measured in units of magnetic diffusion time $t_m = t/Pm$. The black dots indicate the reference data taken from \citet{Vantieghem2016}.}
	\label{fig:Emag}
\end{figure}
The numerical tests of the case P8 are run on grid G48, G64, G80 and G96 as well until $t=8$, with the magnetic time $t_m = t/Pm = 1$.
The time step sizes corresponding to these grid levels are given in Table \ref{tab:gridlevels}.
The time evolution of the magnetic energy $E_{\mathrm{mag}}$ on the grid G64 is displayed in Fig. \ref{fig:Emag}, which also shows an oscillating solution obtained by decreasing the initial magnetic field $\mathbf{B}'_{\mathrm{initial}} = \mathbf{B}_{\mathrm{initial}} / \sqrt{2}$. 
We can see from the figure that the magnetic energy $E_{\mathrm{mag}}$ reaches a constant value within one magnetic diffusion time for the quasi-steady solution while the oscillating solution finally exhibits oscillation behaviour.
And it is clear that the time evolution of the magnetic energy are quite consistent with the benchmark \citep{Vantieghem2016}.

\begin{table}
	\caption{Comparison with the benchmark results for the case P8. G48, G64, G80 and G96 are our numerical results in four different spatial resolutions. FV64 and FV128 refer to the finite volume results in \citet{Vantieghem2016} with six blocks of $64^3$ and $128^3$ grid points, respectively. PS denotes the suggested benchmark solution \citep{Vantieghem2016} obtained by the pseudospectral method. The values in parentheses are relative errors compared with the PS results.}
	\label{tab:comparison}
	\centering
	\footnotesize
	\begin{tabular}{llllllll}
		\toprule
		Results & $R$  & $E_{\mathrm{mag}}$ & $E_{\mathrm{kin}}$ & $T'$ & $u'_\phi$ & $B'_\theta$ & $\omega$ \\
		\midrule
         	G48   & 100  &  313663.1(0.29\%) & 21928.4(1.36\%) & 0.39391(0.18\%) & $-80.08(1.05\%)$ & 2.2556(3.36\%) & 5.0082(9.91\%) \\
		G64   & 133  &  312804.6(0.02\%) & 21758.5(0.57\%) & 0.39345(0.06\%) & $-80.62(0.39\%)$ & 2.2099(1.26\%) & 5.3562(3.65\%) \\
		G80   & 166  &  312664.3(0.03\%) & 21713.5(0.36\%) & 0.39336(0.04\%) & $-80.78(0.18\%)$ & 2.1974(0.69\%) & 5.4563(1.84\%) \\
		G96   & 200  &  312603.7(0.05\%) & 21683.7(0.23\%) & 0.39328(0.02\%) & $-80.86(0.09\%)$ & 2.1905(0.38\%) & 5.5115(0.85\%) \\
		\\
		FV64  & 116  & 309086.0(1.17\%) & 21502.1(0.61\%) & 0.3920(0.31\%) & $-81.23(0.36\%)$ & 2.1510(1.43\%) & 5.6453(1.55\%) \\
		FV128 & 233  & 311950.2(0.26\%) & 21576.2(0.27\%) & 0.3925(0.18\%) & $-80.74(0.24\%)$ & 2.1839(0.07\%) & 5.4959(1.13\%) \\
		PS    &                        & 312754.7 & 21634.9 & 0.3932 &  $-80.9318$ & 2.1823 & 5.5588 \\
		\bottomrule
	\end{tabular}
\end{table}

\begin{figure}
	\centering
	\subfloat[]{\includegraphics[width=0.5\textwidth]{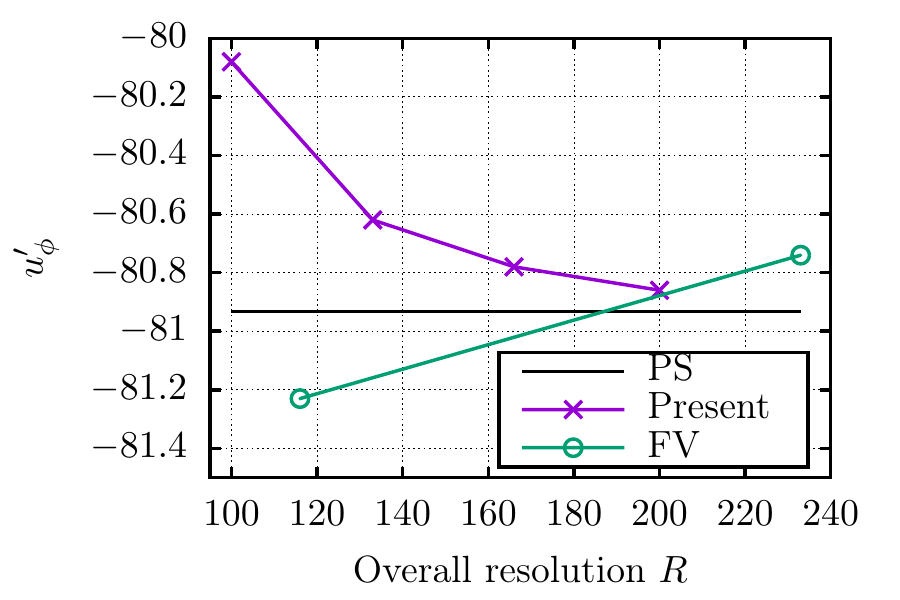}}
	\subfloat[]{\includegraphics[width=0.5\textwidth]{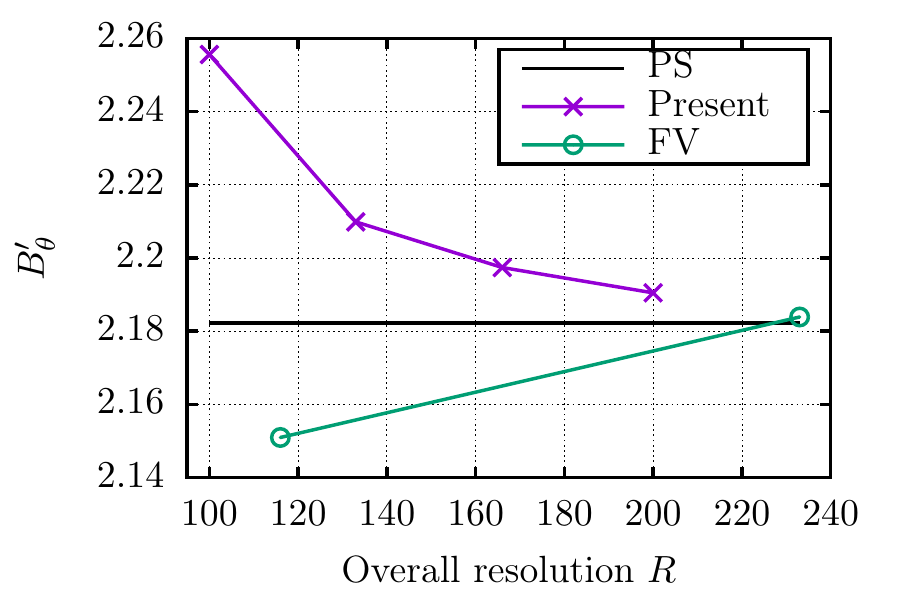}}
	\\
	\subfloat[]{\includegraphics[width=0.5\textwidth]{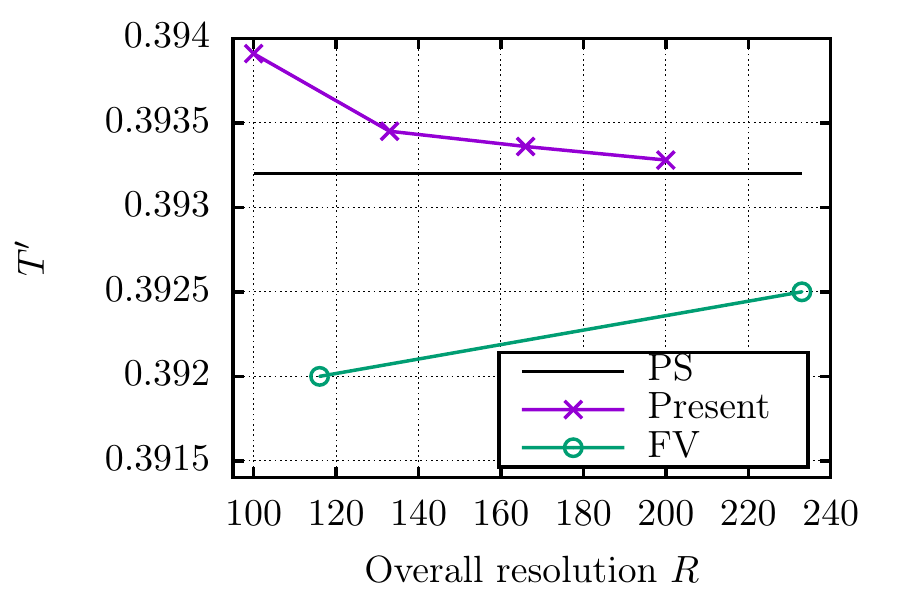}}
	\subfloat[]{\includegraphics[width=0.5\textwidth]{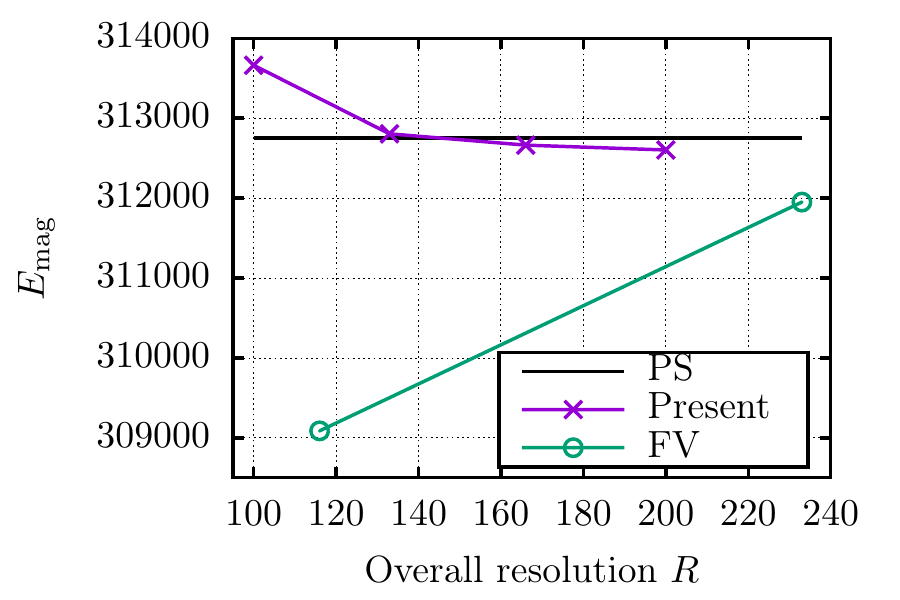}}
	\\
	\subfloat[]{\includegraphics[width=0.5\textwidth]{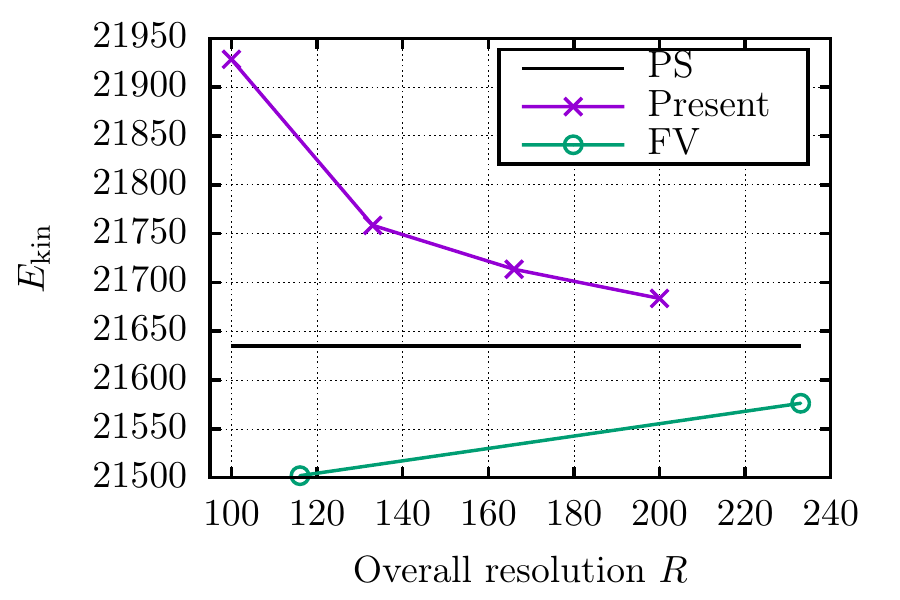}}
	\subfloat[]{\includegraphics[width=0.5\textwidth]{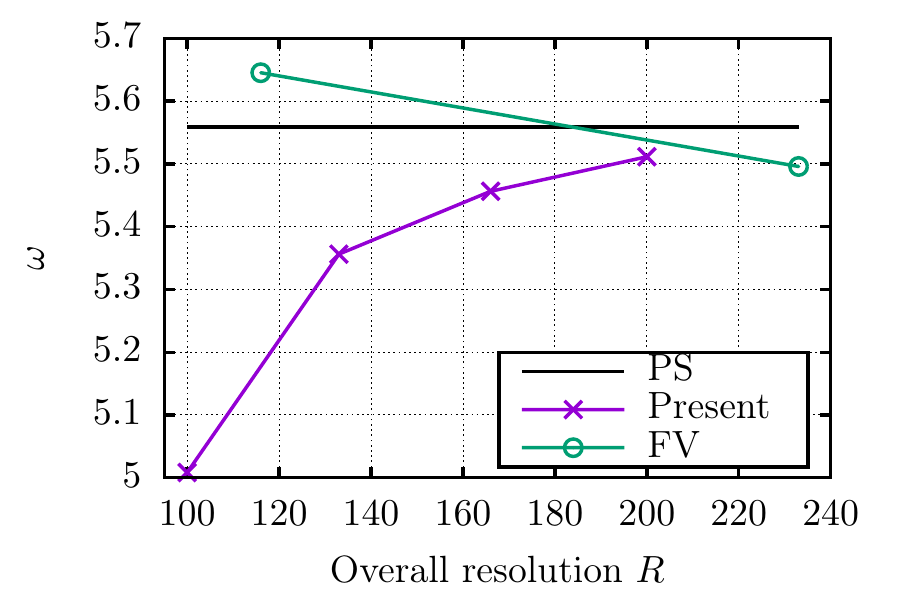}}
	\caption{Convergences as a function of the overall resolution $R$ for the case P8. (a) $u'_\phi$, (b) $B'_\theta$, (c) $T'$, (d) $E_{\mathrm{mag}}$, (e) $E_{\mathrm{kin}}$, (f) $\omega$. The present result and the suggested solution by the pseudospectral method are denoted by Present and PS, respectively. FV refers to the finite volume result reported in \citet{Vantieghem2016}.}
	\label{fig:convergencePm8}
\end{figure}

We calculate the average reference quantities including $E_{\mathrm{mag}}$, $E_{\mathrm{kin}}$, $T'$, $u'_\phi$, $B'_\theta$ and $\omega$ from the final quasi-steady solution and summarize the comparison with the benchmark results in Table \ref{tab:comparison}.
The values in parentheses denote relative errors compared with the benchmark solution obtained by a pseudospectral method.
It can be seen from Table \ref{tab:comparison} that our results are in good agreement with the benchmark pseudospectral solution and the accuracy is comparable to the existing finite volume results.
Besides, the relative errors of the global and local quantities become smaller as the spatial resolution increases.
We display the convergences of these quantities as a function of the overall resolution $R$ in Fig. \ref{fig:convergencePm8}, where the finite volume result reported in \citet{Vantieghem2016} is also shown for comparison.

\begin{figure}
	\centering
	\subfloat[$T'$]{\includegraphics[width=0.48\textwidth]{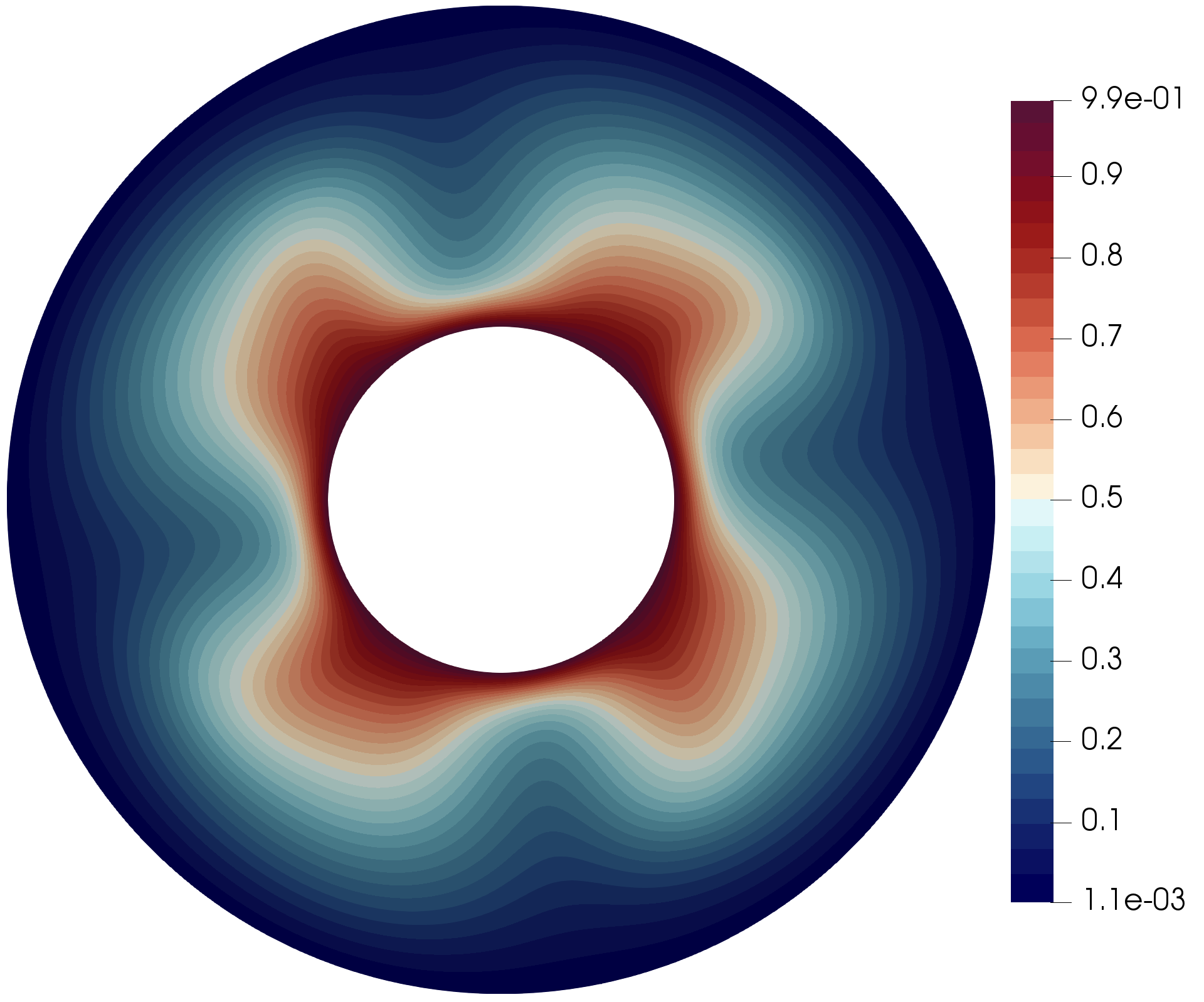}}
	\hfill
	\subfloat[$B'_\theta$]{\includegraphics[width=0.48\textwidth]{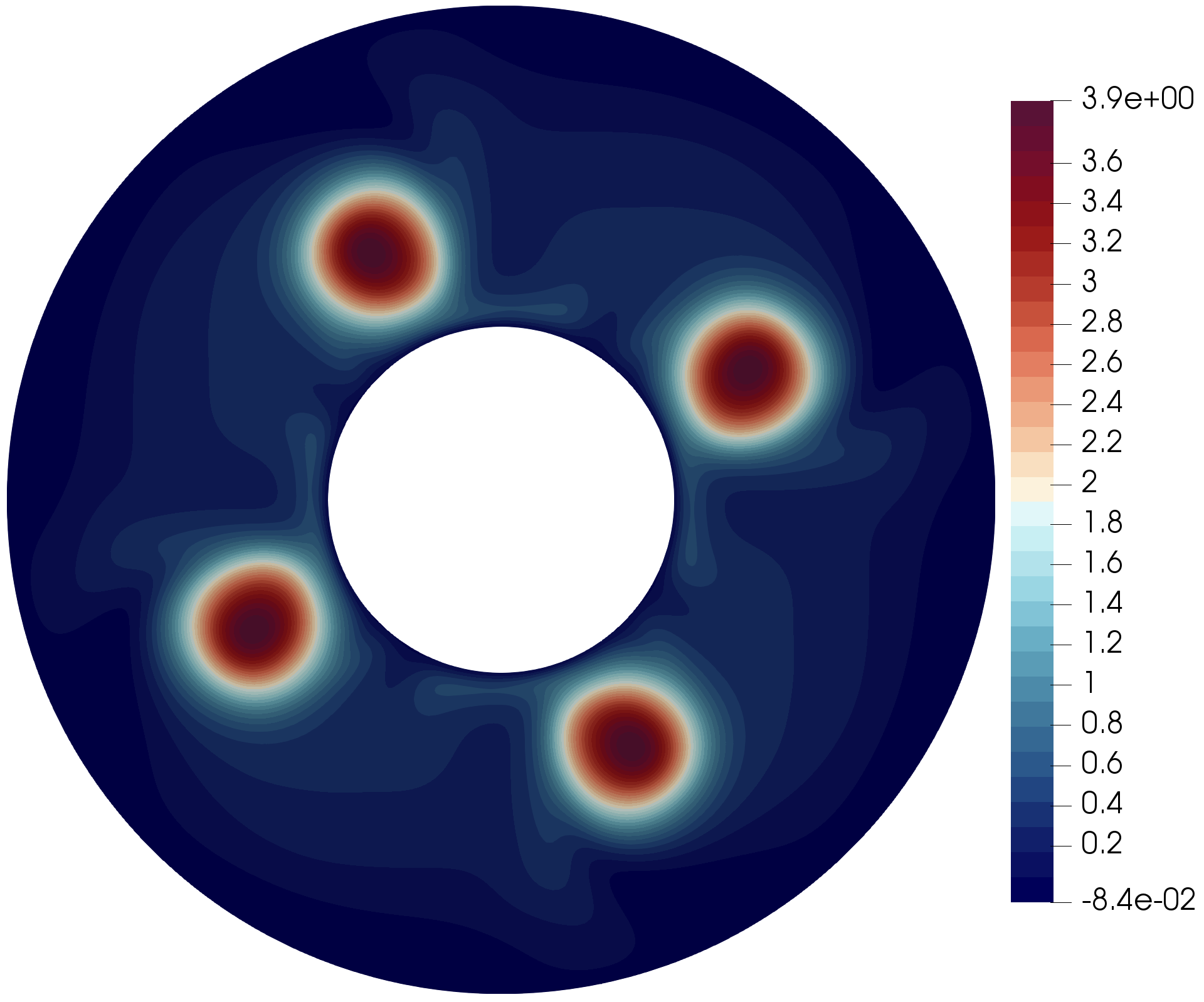}}
	\\
	\subfloat[$u'_\phi$]{\includegraphics[width=0.48\textwidth]{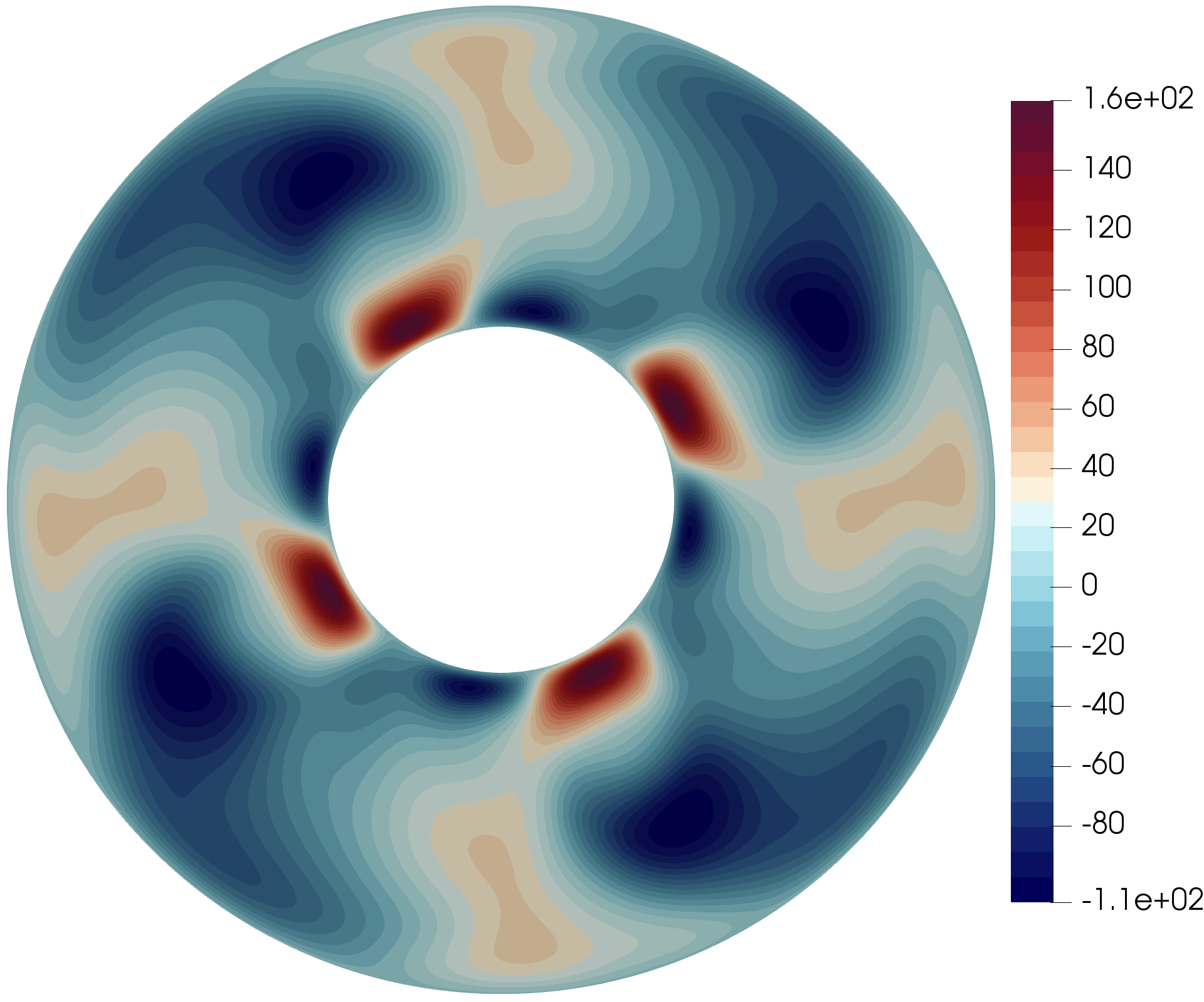}}
	\hfill
	\subfloat[$u'_r$]{\includegraphics[width=0.48\textwidth]{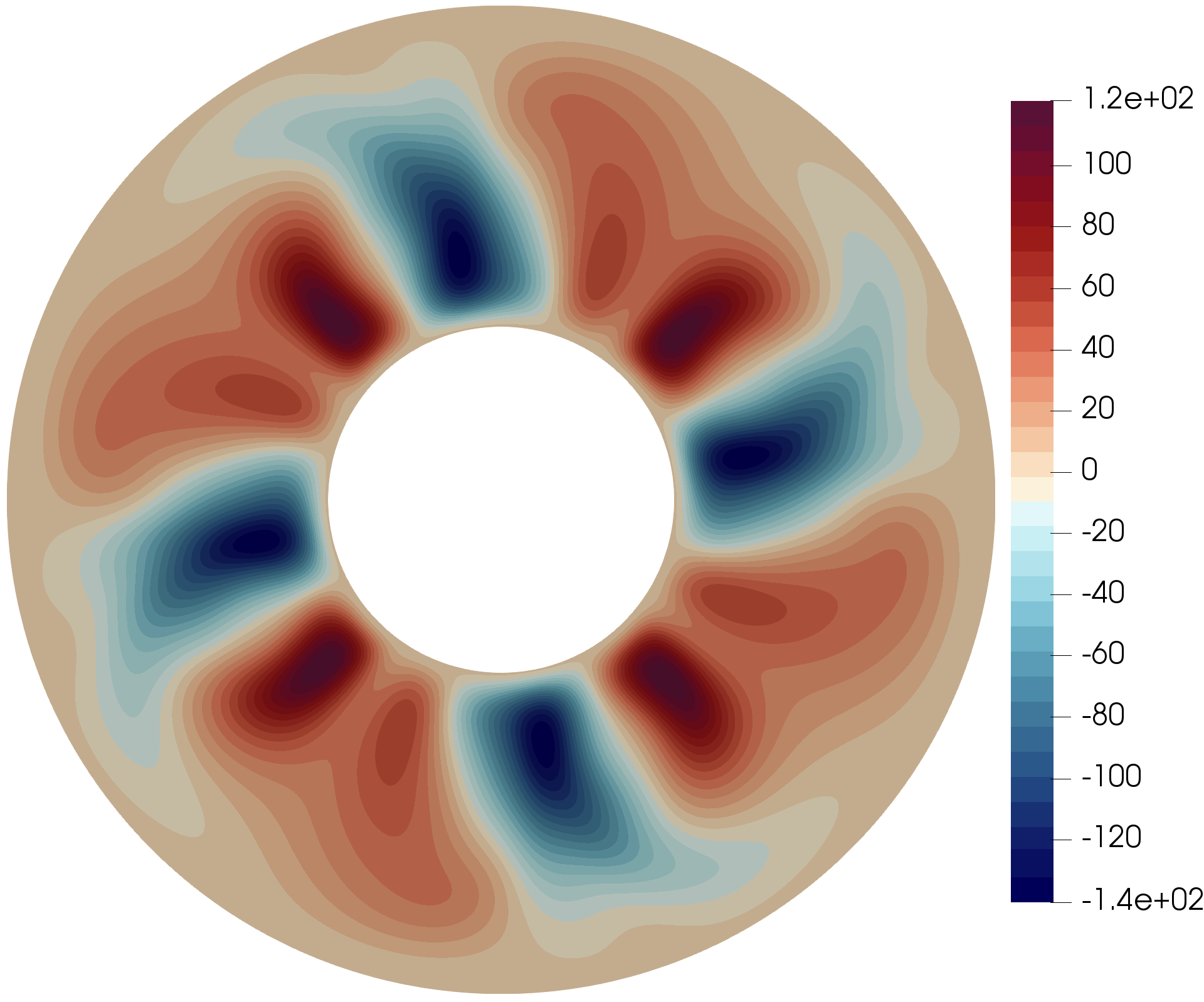}}
	\caption{Equatorial slices of the quasi-steady solution on the grid G96 including $T'$ (a), $B'_\theta$ (b), $u'_\phi$ (c) and $u'_r$ (d).}
	\label{fig:equator}
\end{figure}

\begin{figure}
	\centering
	\subfloat[$T'$]{\includegraphics[width=0.9\textwidth]{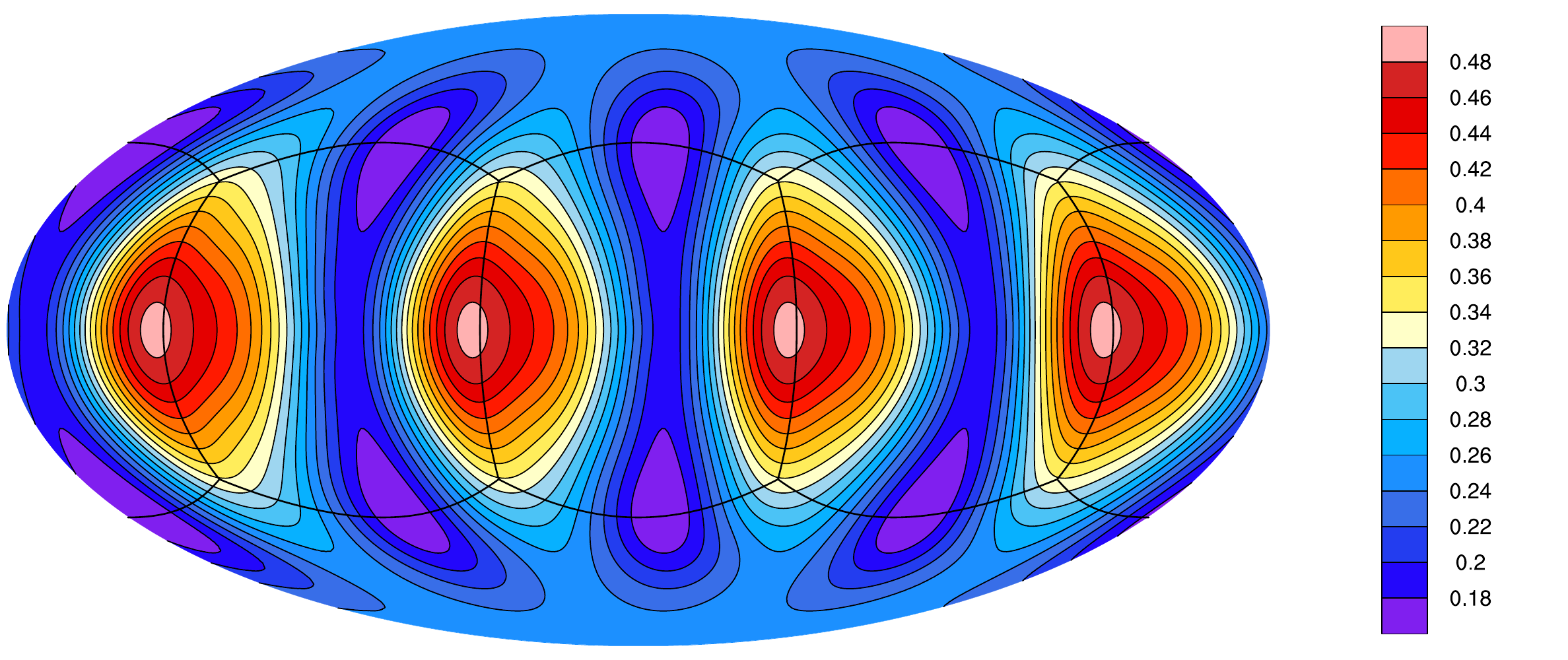}} \\
	\subfloat[$u'_r$]{\includegraphics[width=0.9\textwidth]{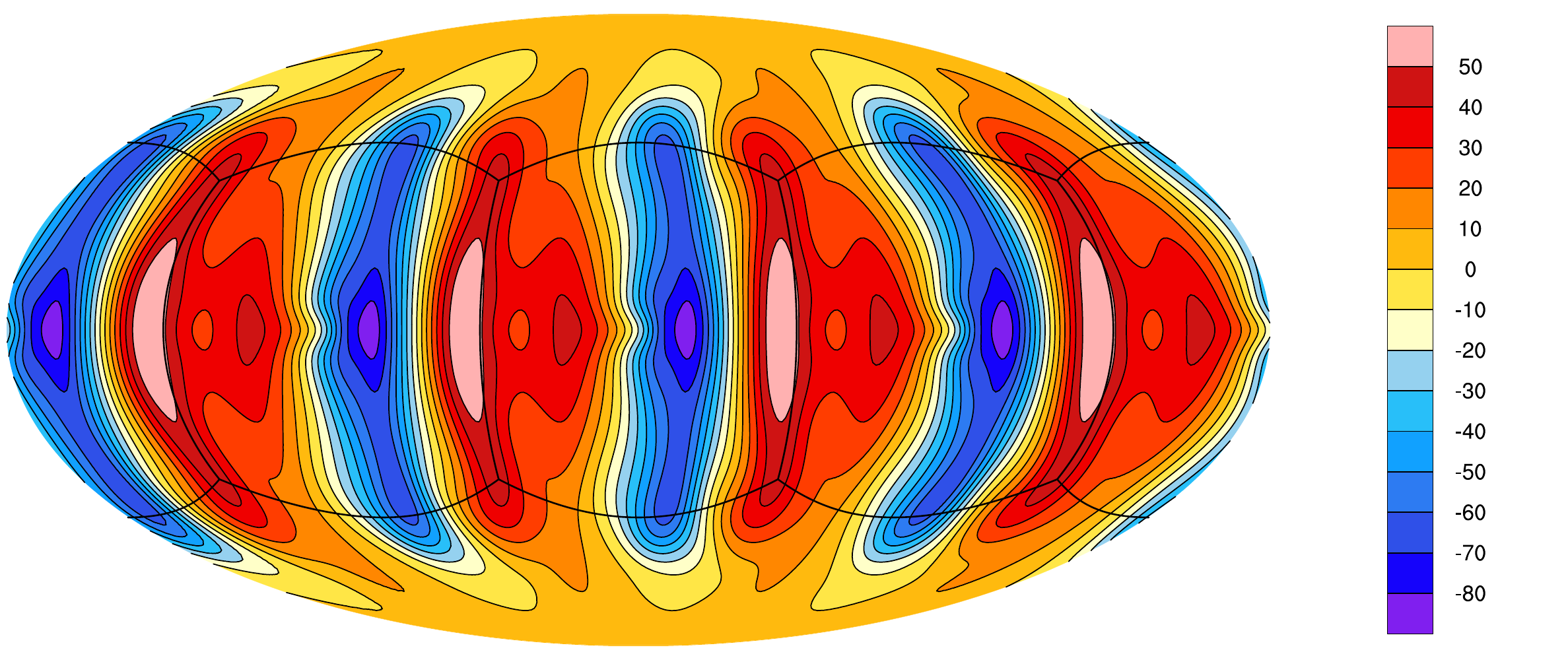}} \\
	\subfloat[$B'_r$]{\includegraphics[width=0.9\textwidth]{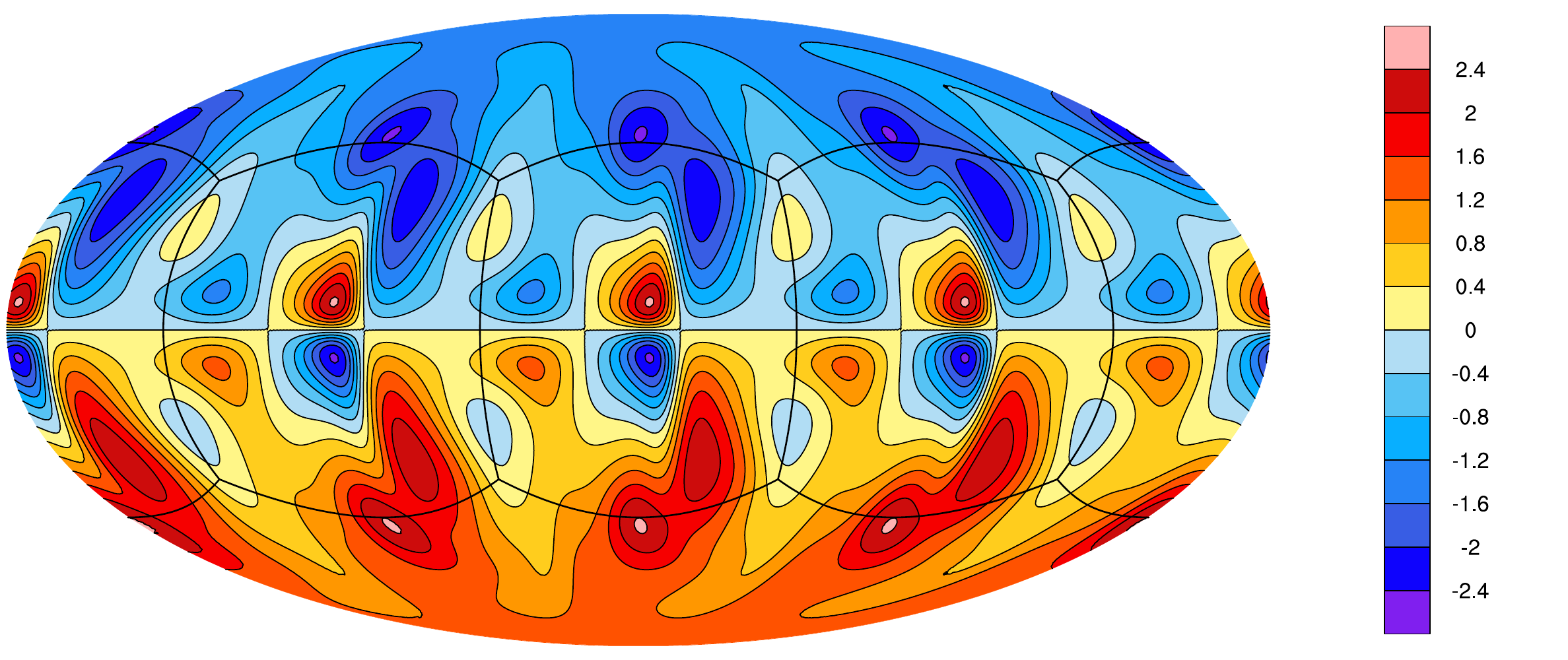}} \\
	\caption{Contours on the mid-depth spherical surface of the quasi-steady solution on the grid G96 including $T'$ (a), $u'_r$ (b) and $B'_r$ (c). The block interfaces of the cubed-sphere grid are denoted by black lines.}
	\label{fig:surface}
\end{figure}

\begin{figure}
	\centering
	\includegraphics[width=0.9\textwidth]{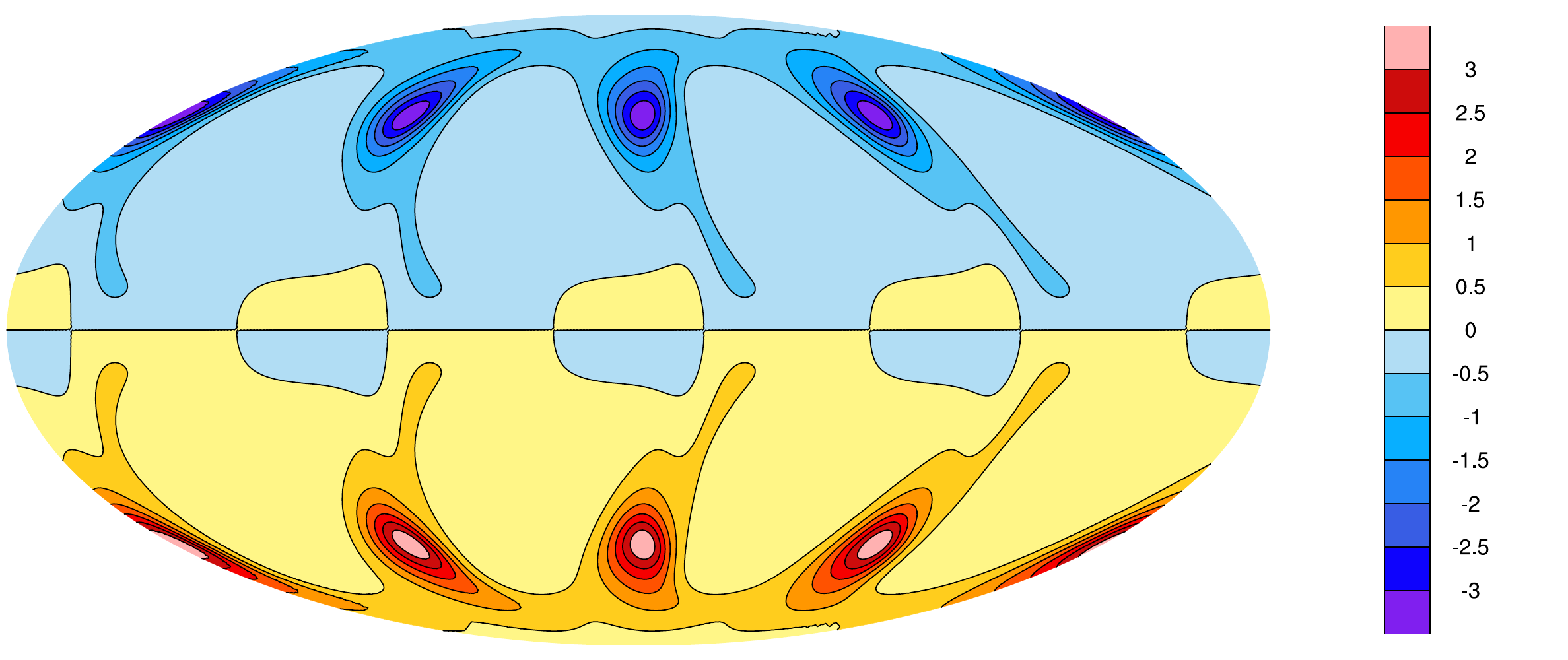}
	\caption{Contour of $B'_r$ on the outer boundary surface of the quasi-steady solution on the grid G96.}
	\label{fig:outer}
\end{figure}

The spatial structure of the quasi-steady solution on the grid G96 is distinctly shown in Figs \ref{fig:equator}--\ref{fig:outer}.
Fig. \ref{fig:equator} depicts the equatorial slices of the quasi-steady quantities including $T'$, $B'_\theta$, $u'_\phi$ and $u'_r$.
It shows good agreement with the benchmark results \citep{Vantieghem2016}.
Fig. \ref{fig:surface} gives the contours on the mid-depth spherical surface of $T'$, $u'_r$ and $B'_r$, and Fig. \ref{fig:outer} displays the contour of $B'_r$ on the outer boundary surface.
The spatial structure of the four-fold azimuthal symmetry can be observed from these figures.

\subsection{Parallel performance}
The parallel performances of the one-level and two-level RAS preconditioner for the case P8 are reported systematically in this section.
In terms of the solver options, the overlap size is $\delta = 1$ and the subdomain solver is ILU(0) for the one-level RAS preconditioner.
For the two-level method, the overlap size $\delta=1$ and the subdomain solver ILU(0) are used for both the fine level and coarse level, while the relative tolerance of the inner GMRES on the coarse level is set to be $0.1$.
We apply both the one-level and two-level preconditioner to the solution of PPBE and compare the two results on efficiency and performance.
The VTBE is only solved by GMRES with the one-level preconditioner.

Firstly, the strong scalability of the GMRES algorithm is studied with a fixed mesh $144 \times 144 \times 168 \times 6$ (about 20.9 million mesh cells) and a constant time step size $\Delta t = 1 \times 10^{-5} (CFL_A = 0.33)$.
The strong scalability refers to the influence of the number of processor cores on the compute time for the problem with a fixed spatial resolution. 
In the ideal situation, the compute time should be reduced proportionally as the number of processor cores increases.
We double the number of processor cores from 1296 up to 10368 and average the corresponding iteration number and compute time of GMRES over the first ten time steps.
The averaged results are summarized in Table \ref{tab:strong}.
\begin{table}
	\caption{Strong scaling results with a fixed mesh $144 \times 144 \times 168 \times 6$ and a constant time step size $\Delta t = 1 \times 10^{-5}$. The results are averaged over the first ten time steps. The averaged iteration numbers of the inner GMRES on the coarse mesh are given in parentheses when using the two-level RAS preconditioner. $np$ denotes the number of processor cores.}
	\label{tab:strong}
	\centering
	\footnotesize
	\begin{tabular}{cccccccccc}
		\toprule
		\multirow{3}{*}{$np$} & \multicolumn{3}{c}{VTBE} & & \multicolumn{5}{c}{PPBE} \\
		\cmidrule{2-4} \cmidrule{6-10}
		& Iteration number & & Compute time (s) & & \multicolumn{2}{c}{Iteration number} & & \multicolumn{2}{c}{Compute time (s)} 
		\\
		\cmidrule{2-2} \cmidrule{4-4} \cmidrule{6-7} \cmidrule{9-10}
		& one-level & & one-level & & one-level & two-level & &  one-level & two-level \\
		\midrule
		1296 & 3.0 & & 6.65 & & 144.9 & 22.0(13.4) & & 29.69 & 15.80 \\
		2592 & 3.0 & & 3.41 & & 154.0 & 22.2(16.0) & & 16.90 & 10.18 \\
		5184 & 3.0 & & 1.77 & & 158.0 & 22.1(14.0) & & 9.60 & 5.72 \\
		10368 & 3.0 & & 0.92 & & 151.1 & 22.4(16.1) & & 5.24 & 4.20 \\
		\bottomrule
	\end{tabular}
\end{table}

\begin{figure}
	\centering
	\subfloat[VTBE and PPBE]{\includegraphics[width=0.49\textwidth]{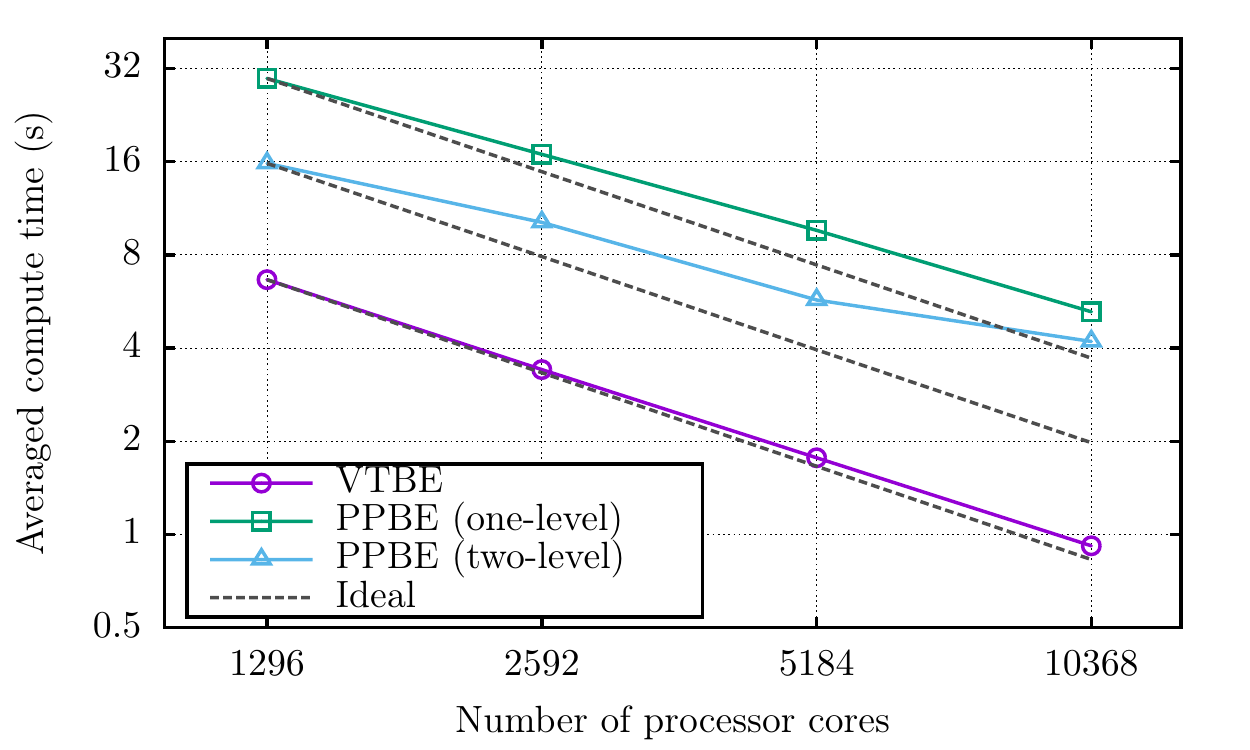} \label{fig:strong_scaling}}
	\subfloat[Breakdown of the two-level method]{\includegraphics[width=0.49\textwidth]{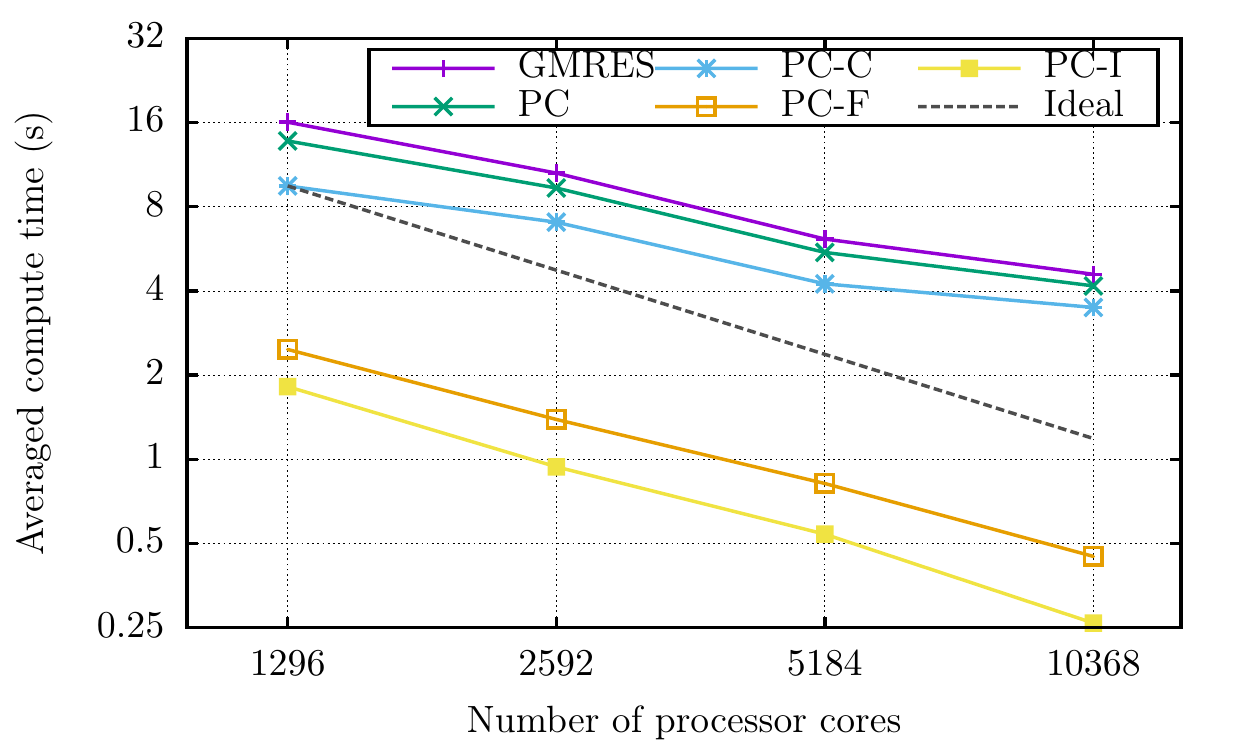} \label{fig:strong_breakdown}}
	\caption{Strong scaling results displaying the averaged compute time with respect to the number of processor cores. (a) The averaged compute time of VTBE and PPBE. (b) Breakdown of the two-level method for PPBE. PC denotes the solution time of the two-level preconditioner in GMRES. PC is then broken down into PC-C (solution time of the linear system on the coarse mesh), PC-F (solution time of the one-level preconditioner on the fine mesh) and PC-I (compute time of the intermediate steps). The dash line refers to the ideal situation.}
	\label{fig:strong}
\end{figure}

It is observed from Table \ref{tab:strong} that the averaged iteration number of GMRES for VTBE remains unchanged as the number of processor cores increases from 1296 to 10368, and the iteration number of PPBE strongly depends on the employed preconditioner.
For the one-level preconditioner, the iteration number of PPBE increases mildly as the number of processor cores is doubled.
With the two-level preconditioner being employed, the iteration number of PPBE is dramatically reduced and is kept to a low level in spite of the double growth in the number of processor cores.
In terms of the compute time of PPBE, the two-level preconditioner is about 20\%--47\% faster than the one-level, which indicates a noticeable improvement of computational efficiency.
Fig. \ref{fig:strong_scaling} displays the averaged compute time of VTBE and PPBE with respect to the number of processor cores.
We can observe from the figure that the GMRES algorithm for VTBE scales very well with up to 10368 processor cores and its strong scalability is quite close to the ideal situation.
The GMRES algorithm for PPBE scales well if the number of processor cores is not too large.
When using a large number of processor cores, e.g. 10368, the strong scalability, as well as the efficiency improvement by the two-level preconditioner, degrade to some extent, because the amount of computations on each processor core is too small.

According to equation \eqref{eqn:coarsesolve} and \eqref{eqn:finesolve}, the solution process of the two-level preconditioner can be broken down into three parts: i) solution of the linear system on the coarse mesh, ii) solution of the one-level preconditioner on the fine mesh, and iii) intermediate steps including restriction, prolongation and others.
The solution times of the two-level method for PPBE divided by these three parts are provided in Fig. \ref{fig:strong_breakdown}.
It is observed that the solution time of the linear system on the coarse mesh is the dominant part of the two-level method.
Specifically, the solution time of the one-level preconditioner on the fine mesh and the time spent on the intermediate steps are quite small and scalable.
Furthermore, the major time-consuming and non-scalable part lies with the linear system solution on the coarse mesh.

To further investigate the performance of the proposed algorithms, we test our code in terms of the weak scalability, which usually draws more interest in practical applications.
The weak scalability focuses on the variation of the compute time with respect to the increase in the number of processor cores while the computational load on each processor core is fixed.
The compute time should remain the same as the number of processor cores grows in the ideal situation.
In our weak scaling test, the time step size is set to be $\Delta t = 1 \times 10^{-5}$ and the mesh size assigned to each processor core is fixed to $20 \times 20 \times 20$.
The number of processor cores is doubled from 648 to 10368 and the corresponding spatial resolution is increased proportionally from $120 \times 120 \times 60 \times 6$ to $240 \times 240 \times 240 \times 6$ (about 82.9 million mesh cells).
As the grid size increases, the value of $CFL_A$ grows from 0.12 to 0.47, which are small enough to obey the numerical stability.
Table \ref{tab:weak} displays the corresponding iteration number and compute time of GMRES for VTBE and PPBE averaged over the first ten time steps.
\begin{table}
	\caption{Weak scaling results with a fixed mesh $20 \times 20 \times 20$ for each processor core and a constant time step size $\Delta t = 1 \times 10^{-5}$. The results are averaged over the first ten time steps. The averaged iteration numbers of the inner GMRES on the coarse mesh are given in parentheses when using the two-level RAS preconditioner. $np$ denotes the number of processor cores.}
	\label{tab:weak}
	\centering
	\footnotesize
	\begin{tabular}{cccccccccc}
		\toprule
		\multirow{3}{*}{$np$} & \multicolumn{3}{c}{VTBE} & & \multicolumn{5}{c}{PPBE} \\
		\cmidrule{2-4} \cmidrule{6-10}
		& Iteration number & & Compute time (s) & & \multicolumn{2}{c}{Iteration number} & & \multicolumn{2}{c}{Compute time (s)} 
		\\
		\cmidrule{2-2} \cmidrule{4-4} \cmidrule{6-7} \cmidrule{9-10}
		& one-level & & one-level & & one-level & two-level & &  one-level & two-level \\
		\midrule
		648 & 3.0 & & 3.36 & & 103.9 & 21.8(10.1) & & 11.02 & 7.48 \\
		1296 & 3.0 & & 3.37 & & 114.1 & 21.7(11.7) & & 12.20 & 7.87 \\
		2592 & 3.0 & & 3.37 & & 162.7 & 24.8(17.0) & & 17.47 & 11.63 \\
		5184 & 4.0 & & 4.78 & & 225.5 & 21.8(16.7) & & 24.56& 10.31 \\
		10368 & 4.1 & & 4.83 & & 240.4 & 21.4(18.9) & & 26.37& 11.47 \\
		\bottomrule
	\end{tabular}
\end{table}

\begin{figure}
	\centering
	\subfloat[VTBE and PPBE]{\includegraphics[width=0.49\textwidth]{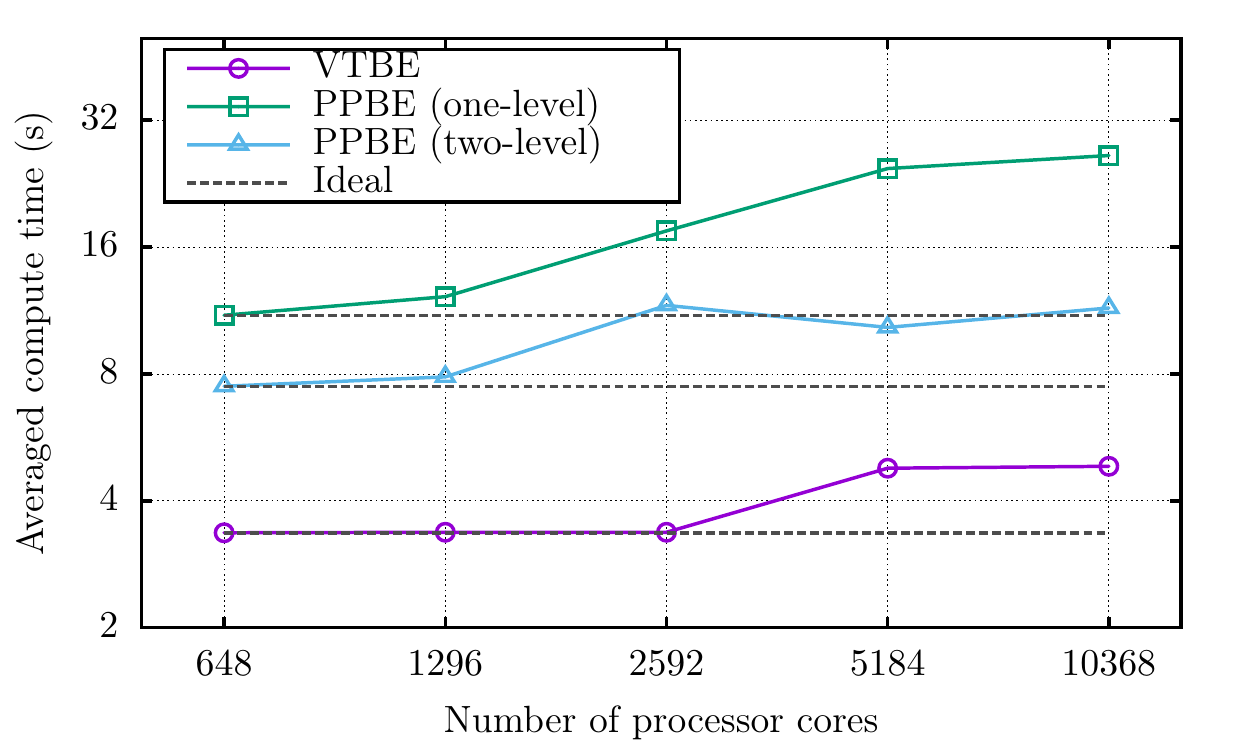} \label{fig:weak_scaling}}
	\subfloat[Breakdown of the two-level method]{\includegraphics[width=0.49\textwidth]{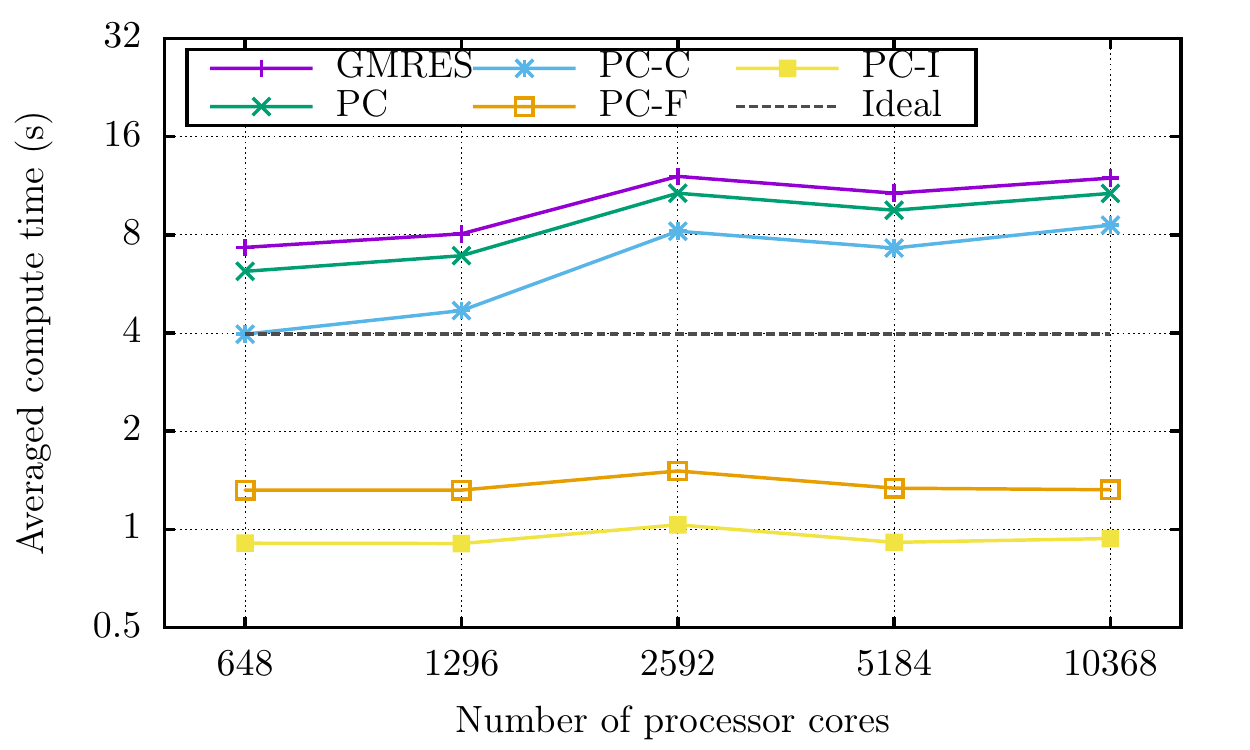} \label{fig:weak_breakdown}}
	\caption{Weak scaling results displaying the average compute time with respect to the number of processor cores. (a) The averaged compute time of VTBE and PPBE. (b) Breakdown of the two-level method for PPBE. PC denotes the solution time of the two-level preconditioner in GMRES. PC is then broken down into PC-C (solution time of the linear system on the coarse mesh), PC-F (solution time of the one-level preconditioner on the fine mesh) and PC-I (compute time of the intermediate steps). The dash line refers to the ideal situation.}
	\label{fig:weak}
\end{figure}

From Table \ref{tab:weak}, we can find that both the iteration number and the compute time of VTBE grow slowly with respect to the increase in the number of processor cores.
The compute time only increases by 44\%, as the number of processor cores increases from 648 to 10368 (16 times larger).
For PPBE, the iteration number of the one-level preconditioner grows fast, while the iteration number of the two-level method stays at a low level.
In terms of the compute time of PPBE, the increase of the one-level preconditioner is 139\% from 648 to 10368 processor cores, which is much larger than 53\% of the two-level method.
In addition, the two-level preconditioner is about 32\%--58\% faster than the one-level as the number of processor cores doubles from 648 to 10368.
The variation of the averaged compute time of VTBE and PPBE with respect to the number of processor cores is further displayed in Fig. \ref{fig:weak_scaling}.
From the figure it can be seen that the VTBE and the PPBE with the two-level RAS preconditioner scales quite well while the PPBE with the one-level method scales a little worse in terms of the weak scalability.
Fig. \ref{fig:weak_breakdown} shows the time breakdown of the two-level preconditioner and it is found that the linear system solution on the coarse mesh is also the dominant part as to the compute time and scalability.
Considering the solution time of the coarse linear system strongly depends on the iteration number of the inner GMRES, which can be easily obtained via multiplying the iteration number on the fine mesh by that on the coarse mesh (e.g. $21.8\times10.1$ for $np=648$), it is understandable that the compute time of PPBE with the two-level method increases with the number of processor cores.

\section{Conclusions}
\label{conclusions}

A scalable parallel solver for the convection-driven magnetohydrodynamic dynamo problem in a rapidly rotating spherical shell with pseudo-vacuum magnetic boundary conditions is developed in this paper.
A finite volume method on a collocated quasi-uniform cubed-sphere grid is employed for the spatial discretization of the spherical shell dynamo equations.
In terms of the temporal integration, a second-order approximate factorization method, applied successfully to the non-magnetic thermal convection problem in our previous study \citep{Yin2017}, is extended to the dynamo governing equations, resulting in two linear algebraic systems, VTBE and PPBE, that are both solved by a preconditioned Krylov subspace iterative method.
To improve the computational efficiency and parallel performance, we design a multi-level restricted additive Schwarz preconditioner based on domain decomposition and multigrid method.
We perform the simulations of two benchmark cases suggested respectively by \citet{Jackson2014} and \citet{Vantieghem2016} and obtain highly accurate numerical solutions, comparable to the existing local method results reported in \citep{Jackson2014, Vantieghem2016, Matsui2016}.
Several numerical tests are carried out to investigate the computational efficiency and the parallel performance with up to 10368 processor cores on the Sunway TaihuLight supercomputer.
The solver of VTBE with the one-level restricted additive Schwarz preconditioner shows very good strong and weak scalabilities.
For the solver of PPBE, a noticeable improvement in the computational efficiency and the weak scalability by the two-level preconditioner is observed, comparing to the one-level method.

To extend our code to the full dynamo problem, the implementations of the insulating boundary condition and the singularity in the inner core should be taken into consideration in the future.
Possible solutions may include an integral boundary element approach \citep{Iskakov2004} together with a parallel fast multipole method \citep{Benson2014} for the issue of the insulating boundary condition and a logically rectangular grid suggested by \citep{Calhoun2008} for the inner core problem.

\section*{Acknowledgements}
We would like to express our appreciation to the anonymous reviewers for their invaluable comments, which have highly improved the quality of the paper.
This work was supported by Beijing Natural Science Foundation, Grant No. JQ18001; by National Natural Science Foundation of China, Grant No. 41174056; by the B-type Strategic Priority Program of the Chinese Academy of Sciences, Grant No. XDB41000000; by the Hong Kong-Macau-Taiwan Cooperation Funding of Shanghai Committee of Science and Technology No. 19590761300.
K. Zhang is supported by the Macau Foundation, by the Pre-research Project on Civil Aerospace Technologies No. D020308 funded by China National Space Administration and by the Macau Science and Technology Development Fund grant No. 0001/2019/A1.

\section*{References}
\bibliography{dynamo_pseudo-vacuum}

\end{document}